\documentclass[lettersize,journal]{IEEEtran}
\usepackage{amsmath,amsfonts}
\usepackage{array}
\usepackage{textcomp}
\usepackage{stfloats}
\usepackage{url}
\usepackage{verbatim}
\usepackage{graphicx}
\usepackage{cite}
\usepackage{hyperref}           
\usepackage{braket}             
\usepackage{amsthm}             
\usepackage{subcaption}

\usepackage{nicematrix} 
\usepackage{tikz}
\usepackage{xcolor}
\usepackage{tikz-network}
\usepackage{adjustbox}
\usepackage{comment}
\usepackage{booktabs}
\usepackage{tikz}
\usepackage{multirow}
\usetikzlibrary{arrows.meta, positioning, calc, shapes.multipart}
\usetikzlibrary{quantikz2}
\usepackage[most]{tcolorbox}
\usepackage{amsmath}
 \usepackage{amssymb}
\usepackage[ruled,vlined,linesnumbered,noend]{algorithm2e}
\SetKwInput{KwIn}{Input}
\SetKwInput{KwOut}{Output}
\SetKw{KwRet}{return}
\SetKwComment{Comment}{$\triangleright$ }{}
\usepackage[mathscr]{euscript} 
\usepackage{booktabs,tabularx,array}

\usetikzlibrary{fit,arrows,matrix}

\definecolor{myBlue}{RGB}{0,51,101}
\definecolor{myNeighborSet}{HTML}{005B5F}
\definecolor{myTrackingSet}{HTML}{735D9F}
\definecolor{myPurple}{RGB}{158,57,106}
\definecolor{darkblue}{HTML}{1f4e79}
\newcounter{boxcounter}

\tcbuselibrary{skins}

\newcolumntype{Y}{>{\raggedright\arraybackslash}X}
\newcolumntype{L}[1]{>{\raggedright\arraybackslash}p{#1}}

\theoremstyle{remark}

\theoremstyle{defin}

\newtheorem{defin}{Definition}

\newtheorem{remark}{Remark}

\begin{document}

\title{A Quantum Internet Protocol Suite\\ Beyond Layering}
\author{Angela Sara Cacciapuoti and Marcello Caleffi\\
    \textsc{Invited Paper}
    \thanks{A.S. Cacciapuoti and M. Caleffi are with the \href{www.quantuminternet.it}{www.QuantumInternet.it} research group, University of Naples Federico II, Naples, 80125 Italy. E-mail: \href{mailto:angelasara.cacciapuoti@unina.it}{angelasara.cacciapuoti@unina.it},  \href{mailto:marcello.caleffi@unina.it}{marcello.caleffi@unina.it}.}
    \thanks{This work has been funded by the European Union under Horizon Europe ERC-CoG grant QNattyNet, n.101169850 (\href{https://qnattynet.quantuminternet.it}{www.QNattyNet.QuantumInternet.it}). Views and opinions expressed are however those of the author(s) only and do not necessarily reflect those of the European Union or the European Research Council Executive Agency. Neither the European Union nor the granting authority can be held responsible for them.}
}

\maketitle

\begin{abstract}
Layering, the organization principle of the network protocols underpinning the classical Internet, is ill-suited to the Quantum Internet, built around entanglement, which is non-local and stateful. This paper proposes a quantum-native organizational principle based on \textit{dynamic composition}, which replaces static layering with a distributed orchestration fabric driven by the node local state and ``in-band'' control. Each node runs a \textbf{Dynamic Kernel} that (i) constructs a local \textit{Plan of Actions} (PoA) of candidate steps to advance a given service intent, and (ii) executes the PoA by composing atomic \textit{micro-protocols} into context-aware procedures, namely the \textit{meta-protocols}. Quantum packets carry an in-band control-field -- the \textit{meta-header} -- containing the service intent and an append-only list of \textit{action-commit records}, termed as \textit{stamps}. Successive nodes exploit this minimal, authoritative history to construct their local PoAs. As quantum packets progress, these local commits collectively induce a network-wide, \textit{direct acyclic graph} that certifies end-to-end service fulfillment, without requiring global synchronization or an external controller. In contrast to classical encapsulation, which prescribes a fixed vertical processing order, the proposed suite enforces order by certification: dependency-aware local scheduling decides what may run at a certain node, stamps certify what did run and constrain subsequent planning. By embedding procedural control within the quantum packet itself, the design ensures coherence and consistency between entanglement-state evolution and control-flow, preventing divergence between resource state ad protocol logic, while remaining MP-agnostic and implementation-decoupled. The resulting suite is modular, adaptable to entanglement dynamics, and scalable. It operates correctly with or without optional control-plane hints. Indeed, when present, hints can steer QoS policies, without changing semantics. We argue that \textit{dynamic composition} is the organizing principle required for a truly quantum-native Internet.
\end{abstract}


\section{Introduction}
\label{sec:01}
The Quantum Internet \cite{CalCac-25,CacCalIll-25,CacCalTaf-20, CacCalVan-20,rfc9340,Kim-08,VanMet-14,DurLamHeu-17} is an entanglement-packet switching network, fundamentally departing from the design principles of the classical Internet~\cite{CalCac-25,CacCalIll-25,CacIllCal-23}. Unlike a classical packet, which carries information from a source to a destination, an \textit{entanglement packet does not transport information}. Its role is to distribute quantum correlations across distant nodes. These correlations can then be subsequently extended, transformed or consumed by local operations performed at any of the involved nodes. 

The ``entanglement-packet'' paradigm \textit{decouples} quantum-state transfer and processing from the physical transmission of carriers: quantum information does not flow through the network. Rather, entanglement -- as the key network resource -- is distributed and later consumed to realize, with the aid of classical communications, distributed quantum functionalities, including information sharing via teleportation, distributed quantum computation and other multi-node state manipulations.

In this light, \textit{the scope of the network} is not to route information, but rather to \textit{distribute, maintain, and manipulate entanglement}. Consequently, while the classical Internet is designed for end-to-end information transfer, the Quantum Internet is designed for \textit{end-to-end entanglement distribution}, enabling applications with no-classical counterpart \cite{rfc9583,WuHuLi-24,LiXueLi-23,LiuAllCai-22,WanRah-22}. 
This shift in \textit{network scope} has profound design implications: entanglement is not a transient data unit, but a shared resource whose value does not reside at a single destination. Its operational utility spans across all nodes that hold it~\cite{IllCalMan-22}, thereby defying the classic source-destination paradigm and making entanglement an inherently stateful, ``network-wide’’ construct.

Once all of this is acknowledged, it becomes clear that protocol behavior and logic must be tightly coupled to the dynamics of entanglement itself. Indeed, as analyzed in~\cite{CalCac-25,CacCalIll-25}, the non-local and stateful nature of entanglement requires to maintain persistent awareness of entanglement-state evolution across all functionalities that, under a classical perspective, would span across the entire protocol stack. Specifically, quantities such as coherence time and fidelity -- at a first glance perceived as physical-layer metrics -- must be visible to functionalities that, in a classical interpretation, would belong to higher layers. For example, entanglement swapping is conceptually associated with a network-layer mechanism, extending connectivity across multiple hops, whereas purification resembles a transport-layer functionality, ensuring the reliability and quality of entangled paths. In this sense, the descriptors of an entanglement resource -- such as fidelity, coherence budget and the identities of the nodes sharing it (ownership) -- are not mere performance indicators, but network \textit{state variables} that drive protocol logic and behavior across all the network functionalities.

These intertwined dependencies cannot be reconciled with a design based on static, vertically isolated layers: any rigid separation of functionalities would fragment state variables that are intrinsically inseparable, by obscuring constraints that must be handled coherently. \textit{In a nutshell, entanglement breaks the very same principle of separation of concerns on which layering design relies}, as it constitutes an intrinsically \textit{cross-layer resource} that shapes and drives the behavior of all network functionalities. Consequently, the dynamics of entanglement must remain visible, shared and actionable across the entire protocol suite.

If one nevertheless insists on a layered design, each layer would then be forced to maintain its own copy of the state variables governing entanglement evolution -- coherence budget, fidelity, and ownership -- in order to operate correctly, while preserving layer independence. Unfortunately, this duplication is possible only by resorting to classical headers and a classical out-of-band control-plane as in~\cite{DiAQiMil22,YooSinKum24,VisHolDia24}, due to the no-cloning constraint of quantum mechanics. This has two critical consequences. First, a layered architecture cannot support quantum-native control~\cite{CalCac-25,CacCalIll-25}, which requires quantum headers and in-band manipulation of quantum states. Second, layering fundamentally introduces severe and unprecedented scalability limitations. Indeed, being entanglement both stateful and non-local, state-dependent information must be consistently available to all nodes sharing a given entanglement resource, potentially across long distances. Tracking how local quantum operations modify fidelity, coherence budget or partner identity via a classical control-plane would therefore require continual classical state synchronization among all involved nodes. The resulting communication overhead grows rapidly and becomes prohibitive even for relatively small entangled clusters. This phenomenon is not unique to quantum networks: even in classical networks, where entanglement is absent, it is known that the number of control messages required per topology change cannot scale better than linearly on Internet-like graphs~\cite{KriClaFal-07}. In the quantum case, however, where the state of an entanglement resource may change with every local quantum operation, the required update rate is dramatically higher. Therefore, a layered architecture -- together with its inevitable reliance on classical headers and out-of-band signaling -- is inherently unsustainable. 

\begin{table*}[t]
    \centering
    \renewcommand{\arraystretch}{1.5}
    \caption{Fundamental Conflicts Between Classical Layering and the Quantum Internet}
    \label{tab:01}
    \begin{tabular}{m{0.22\linewidth} m{0.35\linewidth} m{0.35\linewidth}}
        \hline
        \hline
        \textbf{Layering Features} &
            \textbf{Layering Design Consequence} &
            \textbf{Quantum Internet Features} \\
        \hline
        \multirow[t]{3}{0.9\linewidth}{\raggedright Vertical and static abstraction boundaries between functionalities}
            & Layers operate independently, by hiding their internal mechanisms and states from the layers above and below.
            & Entanglement is non-local, stateful, and inherently cross-layer. Its descriptors -- such as fidelity, coherence time, and partner identities -- directly affect all protocol functions.\\
        \cline{2-3}
            & Each layer assumes the availability of its own local descriptors to drive protocol logic.
            & Entanglement descriptors are global \textit{state variables} that cannot be confined within a single layer without loss of correctness or consistency.\\
        \cline{2-3}
            & Hindered inter-layer dependencies force duplication of information and/or functionalities across layers to preserve layer autonomy.
            & Duplication is incompatible with quantum-native functioning and quantum header: cross-layer visibility can only be achieved via classical headers and a classical out-of-band control-plane.\\
        \hline
        Inherently classical control-plane
            & Control messages required per topology change cannot scale better than linearly on Internet-like graphs.
            & A purely classical control-plane is unscalable: entanglement is non-local, so each local quantum operation altered a shared state, requiring global updates that incur prohibitive overhead and latency even for relatively small entangled clusters.\\
        \hline
        \hline
    \end{tabular}
\end{table*}
These observations -- summarized in Table~\ref{tab:01} -- reveal that layering cannot serve as the organizational principle for a Quantum Internet protocol suite. A new approach is required, one that natively integrates entanglement-state evolution into protocol behavior, without relying on static abstraction boundaries, as described in the following.

\subsection{Contributions}
\label{sec:1.1}
We propose a quantum-native organizational principle based on \textit{dynamic composition} of atomic functionalities, referred to as \textit{micro-protocols (MPs)}, driven solely by local state and in-band procedural control. Specifically, each node runs a \textit{Dynamic Kernel} that:
\begin{itemize}
    \item [(i)] constructs a local \textit{Plan of Actions} (PoA) as a directed acyclic graph (DAG) of candidate actions and their dependencies to advance a given \textit{service intent} (e.g., teleportation);
    \item[(ii)] executes the subset of actions that is currently feasible, given the node internal capabilities and policy constraints, by composing MPs into context-aware procedures, termed \textit{meta-protocols (MePs)}.
\end{itemize}
The PoA is speculative and local: it reflects the node internal state, including classical and quantum forwarding tables, coherence budgets, and optional hints from the control-plane via the Entanglement Distribution Controller (EDC) \cite{CalCac-25,CacCalIll-25}. 

A key design choice is that procedural control of the service is carried ``in-band''. Specifically, quantum packets carry a control-field -- called \textit{meta-header} and orthogonally encoded with respect to the quantum payload -- containing the service intent and an append-only list of \textit{action-commit records}, termed as \textit{stamps}. Successive nodes exploit this minimal, authoritative history to construct their local PoAs, execute feasible actions, and append new stamps at \textit{action-commit boundaries}. By embedding procedural control within the quantum packet itself -- thus, named in-band control -- the design ensures tight coherence between entanglement-state evolution and control-flow and avoids the scalability limits of global, purely classical procedural control.

\begin{figure*}
    \hfill
    \begin{minipage}[c]{0.3\textwidth}
    \centering
        \includegraphics[width=\linewidth]{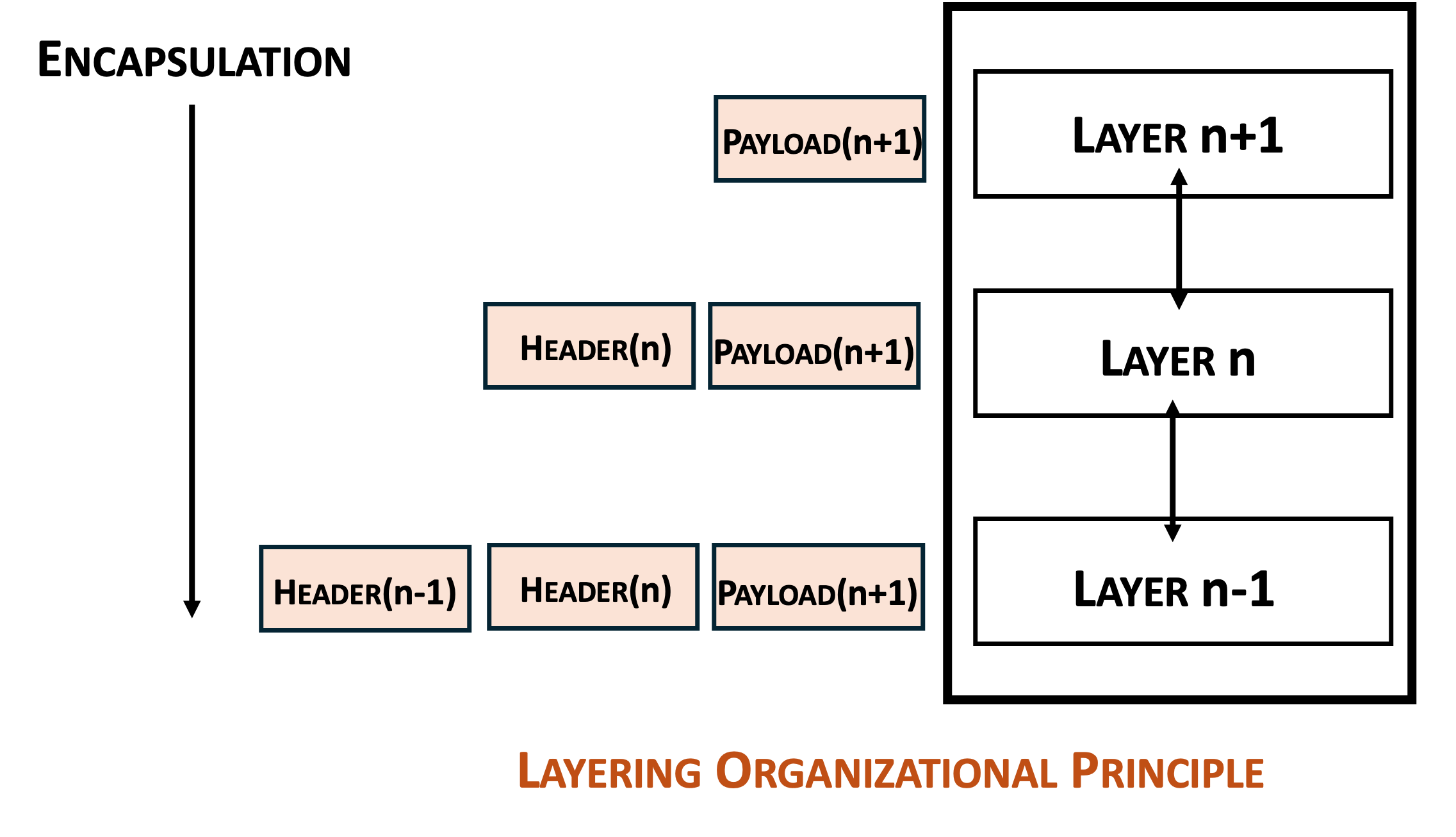}
        \label{fig:1.1}
    \end{minipage}
    \hspace{0.015\textwidth}
    \begin{minipage}[c]{0.02\textwidth}
     \centering
        \tikz{\draw[dashed] (0,0) -- (0,4.2);} 
    \end{minipage}
	\begin{minipage}[c]{0.635\textwidth}
        \centering
        \includegraphics[width=\linewidth]{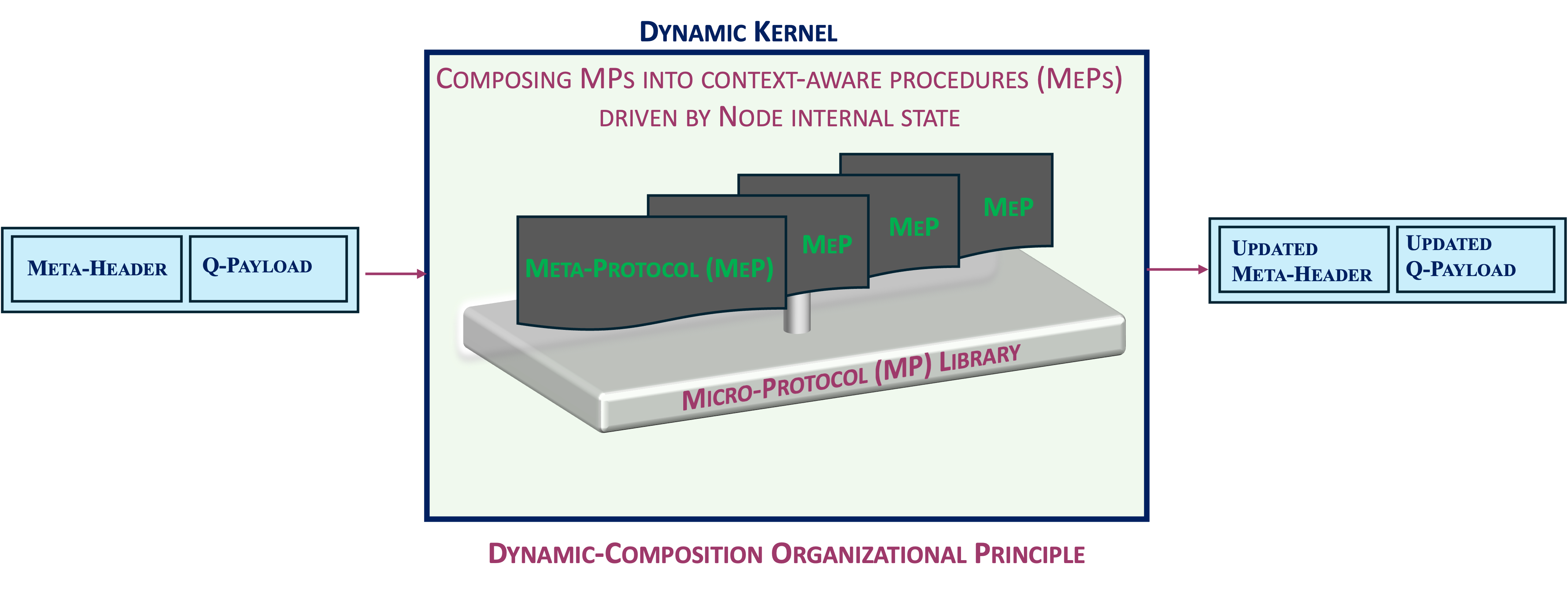}
        \label{fig:1.2}
    \end{minipage}
    \caption{Conceptual shift in the protocol-organizational principle: from static layering to dynamic composition. Each node runs a \textit{Dynamic Kernel} that, based on the node internal state and the in-band meta-header, composes atomic micro-protocols (MPs) into context-aware meta-protocols (MePs) to advance the service intent.}
    \label{fig:01}
	\hrulefill
\end{figure*}

\begin{figure*}
    \centering
        \includegraphics[width=\linewidth]{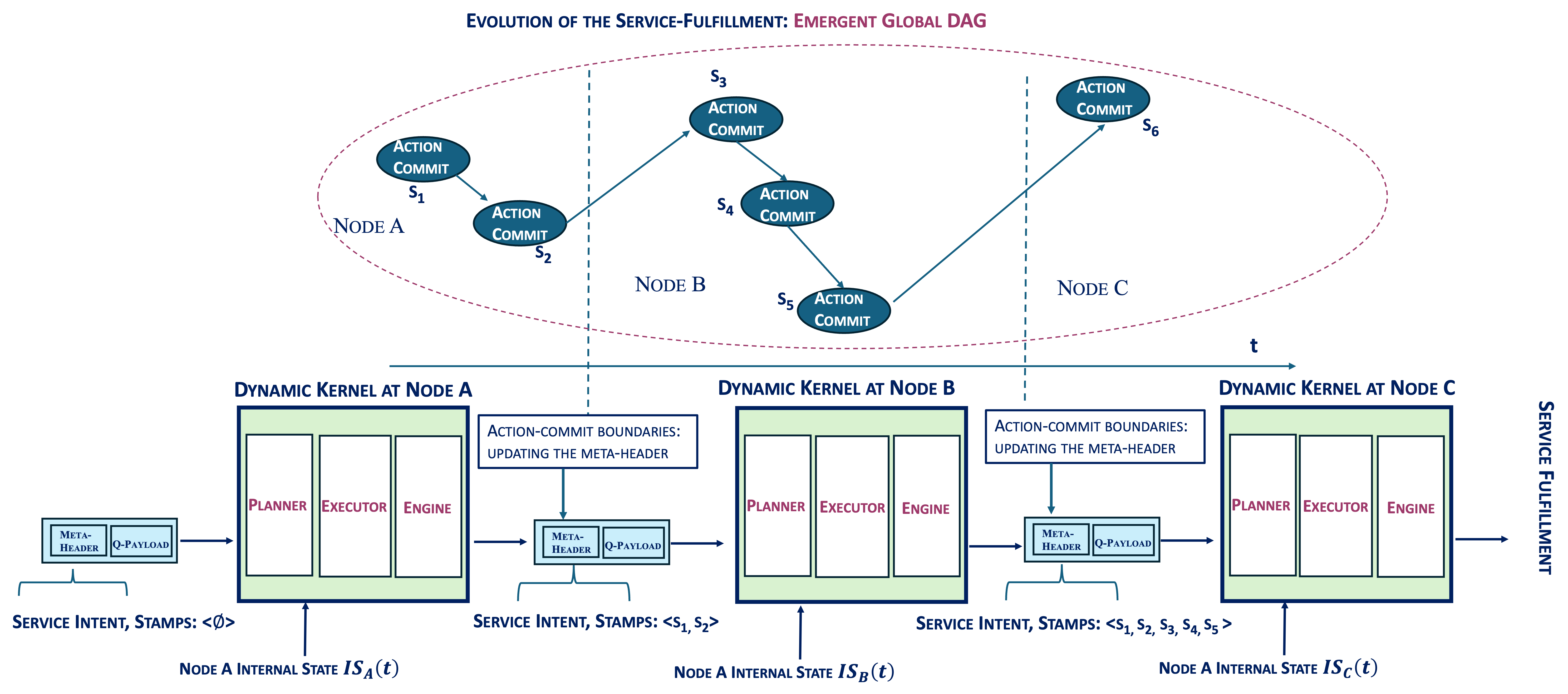}
    \caption{A quantum packet carries a control-field -- called \textit{meta-header} -- containing the service intent and an append-only list of \textit{action-commit records} (\textit{stamps}). Successive quantum nodes exploit this minimal, authoritative history to construct their local PoAs, to execute feasible actions, and append new stamps only at \textit{action-commit boundaries}. Embedding procedural control within the quantum packet itself keeps control-flow aligned with entanglement-state evolution. As packets progress, the accumulated stamps collectively induce a network-wide DAG that certifies end-to-end service fulfillment.}
    \label{fig:02}
	\hrulefill
\end{figure*}

These local commits collectively induce a network-wide DAG that certifies end-to-end service fulfillment, without requiring global synchronization or an external controller. Thus, although each node acts autonomously, the collective effect of local decisions is globally consistent: monotone stamps guarantee an acyclic and verifiable execution history. The pictorial representation of the proposed organization principle is provided in Figs.~\ref{fig:01} and \ref{fig:02}.

Crucially, unlike classical encapsulation, which prescribes a fixed processing order ruled by the vertical hierarchy among layers, the proposed suite enforces order by \emph{certification}, not prescription. More precisely, locally at each quantum node, the PoA DAG and execution scheduling honor dependency edges and resource/quality guards. Globally across the network, the monotone sequence of in-band stamps forms the authoritative history that subsequent nodes must respect. The design is (i) \textit{MP-agnostic} -- thus decoupling mechanism from implementation -- (ii) \textit{modular} and (iii) \textit{adaptive} to entanglement dynamics. It scales, since it avoids duplicated states and out-of-band orchestration. 

It is important to clarify that the \textit{in-band, per-packet procedural control} encoded within the meta-header must not be conflated with any \textit{external} control-plane \cite{CacCalIll-25,CalCac-25}. Analogous to classical protocol headers -- which carry per-packet control information and are distinct from SDN-style out-of-band control -- the meta-header governs local planning/execution and commit certification on a per-packet basis, whereas an external control-plane via EDC may optionally supply hints\footnote{Throughout this paper, ``in-band'' and ``out-of-band'' do not refer to the physical transmission medium but rather to whether or not the procedural control resides in the packet meta-header that carries the service intent. Accordingly, a control message is \textit{out-of-band} whenever it is not carried in the packet meta-header, irrespective of whether it traverses classical or quantum channels.}. When present, control-plane hints are non-intrusive: they prune conservative dependencies and enable QoS policies (e.g., prioritization of actions) without altering semantics. When absent, the suite remains correct. 

In a nutshell, adaptability, modularity, and service progression do not emerge from fixed vertical boundaries but rather from the composability of atomic functions and in-band per-packet procedural control.

By summarizing, the contributions of this paper are threefold:
\begin{enumerate}
    \item It introduces dynamic composition as a quantum-native organizational principle, by replacing layering with a dynamic kernel that composes MPs into context-aware procedures, i.e., the MePs.
    \item It formalizes service fulfillment as an emergent, network-wide DAG of action commits: local kernels collectively construct a consistent end-to-end execution order without centralized control.
    \item It introduces an in-band meta-header, which carries a monotone sequence of action-commit stamps, providing scalable and verifiable coordination for quantum-stateful services, while remaining MP-agnostic and implementation-decoupled. Optional EDC hints can enforce QoS without altering the core semantics of the suite.
\end{enumerate}

The remainder of the paper is organized as follows. Section~\ref{sec:02} introduces some preliminaries utilized through the paper. Section~\ref{sec:03} presents the Dynamic Kernel suite and its MP/MeP composition model. Section~\ref{sec:04} formalizes service fulfillment as an emergent, network-wide DAG of action commits, and it illustrates the suite behavior with an end-to-end teleportation case study. Section~\ref{sec:05} discusses the design implications of the proposed suite and details how it overcomes the limitations of layering-based approaches. Section~\ref{sec:05} also analyzes future research directions. Finally, Section~\ref{sec:06} concludes the paper.

\subsection{Related Work}
\label{sec:1.2}

Layering, the long-standing organizational principle of the classical Internet, achieved modularity and interoperability by enforcing strict vertical and static abstraction boundaries between functionalities. However, it also introduced fundamental limitations: rigidity, duplicated control logic across layers, and hindered the expression of inter-layer dependencies. These limitations are neither new nor subtle in classical networks: they have been very well-recognized in the classical literature \cite{BraFabHan-03,GazPatAlo-10,KreRamVer-14}.

As noted in \cite{BraFabHan-03}, reluctance to change deployed implementations and ossified inter-layer interfaces have repeatedly led classical designers to insert new functionality between existing layers -- ``middle-layer accretions'' -- rather than modify the layers themselves. In \cite{BraFabHan-03}, the authors proposed a Role-Based Architecture (RBA), which abandons strict layering and organizes functionality into ``roles'' with role-specific headers carried in a metadata ``heap''. Ordering and access are governed by explicit processing rules rather than by encapsulation. In a similar spirit, dynamically adaptable protocol stacks \cite{GazPatAlo-10}, cross-layer design, and Software-Defined Networking (SDN) \cite{KreRamVer-14} emerged in the classical domain to mitigate the inefficiencies of rigid layering, allowing dynamic interactions across protocol boundaries and -- via SDN control/data plane decoupling -- programmable and policy-driven composition of network functions. In parallel, John Day's pioneering work \cite{Day-08} (the RINA perspective) re-framed a ``layer'' as a scope-bounded, repeating structure -- a Distributed IPC Facility (DIF) -- rather than a fixed functional stratum, correcting a long-standing misinterpretation of layering traceable to the CYCLADES project led by Louis Pouzin.

In classical networks, the aforementioned shortcomings have often been tolerable because protocol behavior is predominantly local: an operation at one node does not induce intrinsic non-local dependencies across distant nodes. Quantum networks differ fundamentally. Entanglement induces non-local correlations, so local choices constrain remote possibilities. Therefore, the limitations and inefficiency of rigid layering become structural obstacles, as analyzed in Sec.~\ref{sec:01} and summarized in Table~\ref{tab:01}. In short, conventional layering is fundamentally incompatible with the operational requirements of the Quantum Internet. This motivates the proposed \textit{dynamic kernel}, which preserves modularity without freezing inter-function dependencies. 

Nevertheless, most quantum network protocol stacks proposed to date continue to mirror classical-layering design \cite{VanLadMun-08,VanTouHor-11,VanTou-13,MatDurVan-19,VanSatBen-21,Weh-19,KozWeh-19,KozDahWeh-20,PomDonWeh-21,PirDur-19,RamPirDur-21}. A comprehensive survey and analysis of these layered quantum-stack architectures is provided in~\cite{IllCalMan-22}. It was indeed in~\cite{IllCalMan-22} that the core argument advanced in this work was first articulated: a layered design is inherently unscalable in the Quantum Internet, although that study did not propose a technical solution to the problem. 

A recent work~\cite{PirMunDur-25} also recognized that classical layering is unscalable for quantum networks and proposes a task-based control framework. However, that design relies on assumptions that make practical deployment challenging. In particular, it assumes each node maintains a \textit{global, continuously refreshed view} of all quantum resources in the network, updated through classical signaling whenever entanglement is generated, consumed, or reallocated. Maintaining such global visibility entails significant communication and latency overheads, and appears difficult to reconcile with the operational realities of quantum networks, where entanglement evolves through local operations and cannot be synchronized network-wide without prohibitive classical cost.
Indeed, the architecture in~\cite{PirMunDur-25} introduces a globally coordinated classical control layer on top of the quantum substrate. It functions as a control-theoretic scheduler whose correctness depends on strong, system-wide consistency assumptions, unlikely to hold in realistic deployments. 
As a result, \cite{PirMunDur-25} does not provide an organizational principle for a protocol suite. Instead, it externalizes coordination to a global (classical) controller. By contrast, our dynamic-composition approach avoids any requirements of global knowledge, being grounded in distributed, node-centric execution model. Each node operates solely on its local view, while packet-carried stamps allow end-to-end correctness to emerge from local decisions, without requiring centralized orchestration or network-wide synchronization.

\section{Preliminaries}
\label{sec:02}
As introduced in Sec.~\ref{sec:01}, the Quantum Internet is an entanglement-packet switching network, whose fundamental \textit{scope} is to \textit{distribute, maintain, and process entanglement}, which serves as the network delivery unit. Applications -- ranging from quantum teleportation to distributed quantum computing -- \textit{operate on top} of such pre-established entanglement resources. In this light, the network fabric provisions entanglement that applications subsequently consume, without requiring the information-bearing quantum state to traverse each hop. This perspective naturally encompasses both bipartite and multipartite entanglement. Indeed, as detailed in \cite{IllCalMan-22,MazCalCac-25,CheIllCac-24}, protocols that remain valid across different persistency classes of multipartite entanglement can be designed by distributing multipartite resources via teleportation over pre-shared bipartite entanglement. As a consequence, entangled pairs serve as foundational operational primitives for entanglement distribution within the network, while the network objective -- not limited to distribution -- encompasses also the maintenance and processing of entanglement resources at any multipartite scales. 

\subsection{Entities, Planes and Internal States}
\label{sec:2.1}

We adopt the architectural framework introduced in \cite{CalCac-25,CacCalIll-25}, which separates the control-plane from the data plane. In this model, the control-plane comprises a classical component for signaling through standard message exchanges, together with a quantum control-plane that
orchestrates entanglement resources via \textit{Entanglement Distribution Controller} (EDC) \cite{CalCac-25,CacCalIll-25}. A key architectural element inherited from \cite{CalCac-25} is the \textit{quantum packet} structure, consisting of a quantum header and a quantum payload. The quantum header carries the contextual information needed for forwarding and resource selection, while the payload contains entanglement qubits (ebits). This quantum-native packet structure provides the foundation for the meta-header defined in the following.

Let $V$ be the set of network nodes. Each node $v$ maintains an \textit{internal state} $\mathcal{IS}_v(t)$ that evolves over time given by:
\begin{equation} 
\label{eq:01}
\mathcal{IS}_v(t)=\big(FT_c^v(t),\,FT_q^v(t),\,\mathrm{Hints}_v(t)\big),
\end{equation}
where
\begin{itemize}
    \item $FT_c^v(t)$ is the classical forwarding table, capturing the node reachability over the physical network graph.
    \item $FT_q^v(t)$ is the quantum forwarding table \cite{CalCac-25}, namely the entanglement inventory, reflecting the reachability on the entanglement overlay graph. For each resource, $FT_q^v(t)$ records the participating nodes via quantum addresses (QAs), together with a resource identifier and associated descriptors such as fidelity $F$ and coherence budget $\tau$.
    \item $\mathrm{Hints}_v(t)$ denotes optional control-plane assistance, provided through EDC-enforced policies, such as QoS policies or admissible parallelism. 
\end{itemize}
Updates to $\mathcal{IS}_v(t)$ may be periodic or event-driven, for instance due to resource arrival, expiration, consumption, or control-plane policy changes. An entanglement resource is \textit{consumable} only if its descriptors meet the service/application requirements at the time of invocation. Because these descriptors directly gate both feasibility and correctness, they are first-class elements of $\mathcal{IS}_v(t)$, rather than mere performance counters. In other words, because these descriptors directly govern protocol behavior and logic across the suite, they must be part of the node internal state.

\subsection{Service Intent, Micro- and Meta-Protocols}
\label{sec:2.2}

We now formalize the abstractions underlying the proposed protocol suite. We begin with the notion of a \textit{service intent}, a declarative specification of \textit{what} the
application requests -- such as for example end-to-end entanglement establishment, teleportation, or entanglement purification -- independent of \textit{how} the network actually realizes it.

\begin{defin}[Service Intent]
    \label{def:01}
    A \textit{service intent} is an application-issued, mechanism-agnostic request that declares the target quantum functionality and the participating nodes. The intent is opaque to the network. It may include additional parameters permitted by enforced policy. Formally,
    \begin{equation}
        \label{eq:02}
        \mathcal{I} \;=\; \big(\textsf{Service},\, \mathcal{P},\, \Theta,\, \Pi\big),
    \end{equation}
    where (i) $\textsf{Service}$ is the identifier of the requested functionality; (ii) $\mathcal{P}$ is the set of participating nodes, expressed as quantum addresses. In case of bipartite services $\mathcal{P}=\{QA_{\mathrm{src}},QA_{\mathrm{dst}}\}$, whereas for multipartite services, $\mathcal{P}=\{QA_{\mathrm{i}}\}_{\mathrm{i}\subseteq V}$; (iii) $\Theta$ collects quantitative targets, such as for example $F_{\min}$, $\tau_{\min}$; (iv) $\Pi$ carries optional policy annotations, such as QoS class or persistency class in the case of multipartite entanglement. The intent does not encode route information, scheduling, or procedures.
\end{defin}

Stemming from the fundamental objective of the Quantum Internet -- distributing, maintaining, and processing entanglement -- we expose a minimal set of micro-protocols (MPs) and their runtime compositions, called meta-protocols (MePs), as defined in the following. 

\begin{defin}[Micro-Protocol (MP)]
    A micro-protocol implements a single, atomic network functionality.
\end{defin}

Examples\footnote{We provide an example of a canonical set of MPs later in Sec.~\ref{sec:03}} of MPs could include a single link-level entanglement generation attempt, a local quantum operation, or a link-level synchronization step.

\begin{defin}[Meta-Protocol (MeP)]
    A meta-protocol is a higher-order construct obtained by dynamically composing a set of MPs to realize a complex \textit{functionality}. 
\end{defin}
Examples of MePs could include entanglement purification, multi-hop entanglement distribution, and similar compound procedures. As it will become evident in the following, the realization of a complex functionality may require retries, adaptation to evolving entanglement availability, and coordination across multiple nodes. By design\footnote{See Sec.~\ref{sec:03} for further details.}, these features are encapsulated within the MeP and do not leak across it: only \emph{action commits} surface externally as stamps. This preserves a clean separation among mechanism (MPs), procedures (MePs), and certification (stamps). The specific composition, execution and certification workflow is detailed in Sec.~\ref{sec:03}.

\begin{defin}[Action]
    An action is an abstraction representing a unit of progress toward fulfilling a service intent. At run time, an action is realized by instantiating one or more MPs and/or MePs, selected according to the node capabilities, policies, and internal state.
\end{defin}

\begin{defin}[Action Commit]
    The action commit is the certified completion of an action instance. An action commits when its execution reaches a terminal outcome (success or explicit failure), and this event is \textit{recorded by emitting a stamp} $s$. Action commit events are irreversible. 
\end{defin}

By design (see Sec.~\ref{sec:03}), a \textit{stamp} is appended to the packet meta-header, formally defined as follows.

\begin{defin}[Meta-Header]
    \label{def:06}
    The \textit{meta-header} $H$ is the in-band control-field carried by the quantum packet to support end-to-end multi-hop service fulfillment. Formally:
    \begin{equation}
        \label{eq:03}
         H = (\mathcal{I},\,\textsf{Stamps})
    \end{equation}
    and it contains:
    \begin{itemize}
        \item the service intent $\mathcal{I}$, defined in Def.~\ref{def:01};
        \item the ordered, append-only sequence of emitted stamps certifying action commits along the path: $\textsf{Stamps}=\langle s_1,\ldots,s_m\rangle$.
    \end{itemize} 
    The meta-header is encoded in a subspace orthogonal to the quantum payload subspace, thus enabling non-perturbing access. Updates occur only at action-commit boundaries, when the next stamp is appended. The meta-header carries no speculative plans. Its structure is agnostic to the specific MP/MeP implementations.
\end{defin}

We note that the orthogonality\textsuperscript{\ref{foot:04}} between the meta-header and payload subspaces can be realized in different ways, as discussed in Sec.~\ref{sec:5.1}. The specific physical realization of the meta-header is therefore a platform-dependent engineering choice. The architectural design proposed in this paper is intentionally \textit{encoding-agnostic}, since a sound organizational principle must remain invariant under technological evolution and heterogeneous hardware implementations. At the architectural level, the meta-header plays a role fundamentally different from that of classical layered headers. Unlike classical headers, which are bound to a specific layer, the meta-header is a cross-cutting control structure shared across all stages of service fulfillment. As detailed in Sec.~\ref{sec:03}, its role is to provide the next hop with a consistent, authoritative context to advance the service fulfillment. By carrying only executed actions (stamps) -- rather than speculative plans -- the meta-header avoids redundancy and ensures that header growth is bounded by the logical progression of the service, thereby supporting scalability. As a consequence of this design, the meta-header constitutes a programmable \textit{procedural-control} element, while remaining logically coupled to entanglement-state evolution through commit-atomic updates performed exclusively at action boundaries.

It is worth noting that all the above definitions are \textit{invariant} to the evolution of MP/MeP granularity. A construct regarded as a MeP today may collapse into a single MP tomorrow, as technology advances. The abstractions introduced here are stable under such shifts, ensuring that the protocol suite remains scalable, modular, and future-proof.

\section{Protocol Suite Design}
\label{sec:03}

\textbf{Core Idea:} Service fulfillment progresses hop-by-hop and it is driven \textit{in-band} in a fully distributed manner, by the processing performed on the quantum packet. In our design, the packet header is extended into a \textit{meta-header}, defined in Def.~\ref{def:06}, which carries (i) the original service intent and (ii) a sequence of stamps certifying actions already committed along the ``service-path''. No speculative plan of the remaining work is carried. 

The architectural element that allows each node to advance a service intent in a scalable and quantum-native manner is referred to as \textit{Dynamic Kernel}. The kernel reads the meta-header orthogonally to the payload and, using the node internal state $\mathcal{IS}_v(t)$, introduced in eq.~\eqref{eq:01}, \textit{constructs} a local, speculative \textit{Plan of Action} (PoA). The PoA is a \textit{planning artifact} strictly local to the node, since it is never propagated and does not ``evolve'' across hops. Instead, the \textit{service fulfillment} evolves as the packet advances. The kernel executes the subset of planned actions that is locally feasible. Whenever an action reaches a terminal outcome, it \textit{commits} and the kernel appends a new stamp to the meta-header. These stamps provide the next hop with a minimal yet authoritative and consistent record of what has already been completed, enabling that node to recompute its own PoA and further advance the service without requiring any global knowledge or network-wide synchronization. 

\begin{remark}
    A salient property of the proposed suite is its ability to operate correctly under \textit{variable levels of control-plane support}. Indeed, by design, each Dynamic Kernel advances a service intent \textit{autonomously}, by using only node-resident information, namely the service intent, the accumulated stamps (carried in the meta-header), and the node internal state $\mathcal{IS}_v(t)$. 
    Crucially, as it will be evident in the next subsections, correctness and progress do not rely on the $\mathrm{Hints}_v(t)$ field contained in $\mathcal{IS}_v(t)$: the suite remains correct when $\mathrm{Hints}_v(t)=\varnothing$ (purely autonomous operation), i.e., in the complete absence of control-plane assistance. When present, $\mathrm{Hints}_v(t)$ provides optional, policy-driven guidance. Such hints can inject a broader perspective than the one available locally, refining decision-making to improve targeted performance metrics, without compromising local autonomy, which remains essential for scalability. The hints are strictly advisory: they do not modify protocol semantics and may be ignored if absent, stale, or inconsistent, in which case the kernel falls back to the conservative, purely autonomous plan. In other words, EDC hints constitute an optional performance-optimization stratum. This design establishes a clear separation between \textit{correctness}, which is ensured entirely through autonomous, in-band mechanisms, and \textit{optimization}, which may optionally exploit control-plane guidance. As a result, the proposal supports a graceful continuum ranging from fully autonomous operation to control-plane-assisted refinement, without compromising scalability or semantic guarantees.
\end{remark}

\subsection{Design Principle}
\label{sec:3.1}

The proposed suite introduces a quantum-native \textit{organizational principle}
based on \textit{dynamic composition}. Instead of arranging functionalities into a stack of hierarchical layers, each node runs a \textit{Dynamic Kernel} that selects and composes atomic \textit{MPs} into transient, context-dependent \textit{MePs} according to its local state, policies, and the service intent. This organizational principle is governed by three invariants:
\begin{itemize}
    \item \textit{Local Autonomy:}  
    Each node constructs its own PoA solely from its internal state $\mathcal{IS}_v(t)$ and the packet meta-header. No global knowledge or network-wide synchronization is required, thereby eliminating scalability bottlenecks.
    \item \textit{In-Band Coordination:}  
    Control information -- the service intent and the append-only log of action-commit stamps -- travels \textit{in-band} within the quantum packet, encoded in the orthogonal subspace of the meta-header for non-perturbing access. This ensures coherence and consistency between data and control evolution, avoids reliance on out-of-band classical signaling, and prevents divergence between resource state and protocol logic.
    \item \textit{Dynamic Composition:}  
    Based on the locally constructed PoA, each node composes the required MPs into context-aware MePs that realize the actions that are currently ``runnable'' at that node. Sequencing, retries, and fallbacks are handled locally within MePs, since only action commits are stamped. Composition adapts to entanglement availability, coherence budget, and node capabilities (optionally refined by EDC hints) without changing semantics.
\end{itemize}

Together, these three invariants replace static layering with a dynamically coordinated, entanglement-aware orchestration fabric. Modularity is not preserved by vertically isolating functions into layers, but rather by composing and reusing atomic functionalities whose semantics remain invariant across the network. Adaptation, correctness, and service progression arise from the distributed execution of dynamic kernels rather than from fixed inter-layer interfaces.

As will become clear in the remainder of this paper, a key \textit{consequence} of these design principles is that the sequence of locally committed actions, carried hop-by-hop as stamps in the meta-header, implicitly induces an end-to-end DAG representing the global service-fulfillment process. Crucially, this global DAG is neither precomputed nor distributed, nor is it explicitly carried in the packet. Instead, it \textit{emerges} naturally from the autonomous decisions of each kernel and from the monotonic accumulation of stamps certifying action commits. As a result, the global execution structure is obtained without global knowledge, while preserving both correctness and scalability.

\begin{figure*}[!t]
    \centering
    \includegraphics[width=7.5in]{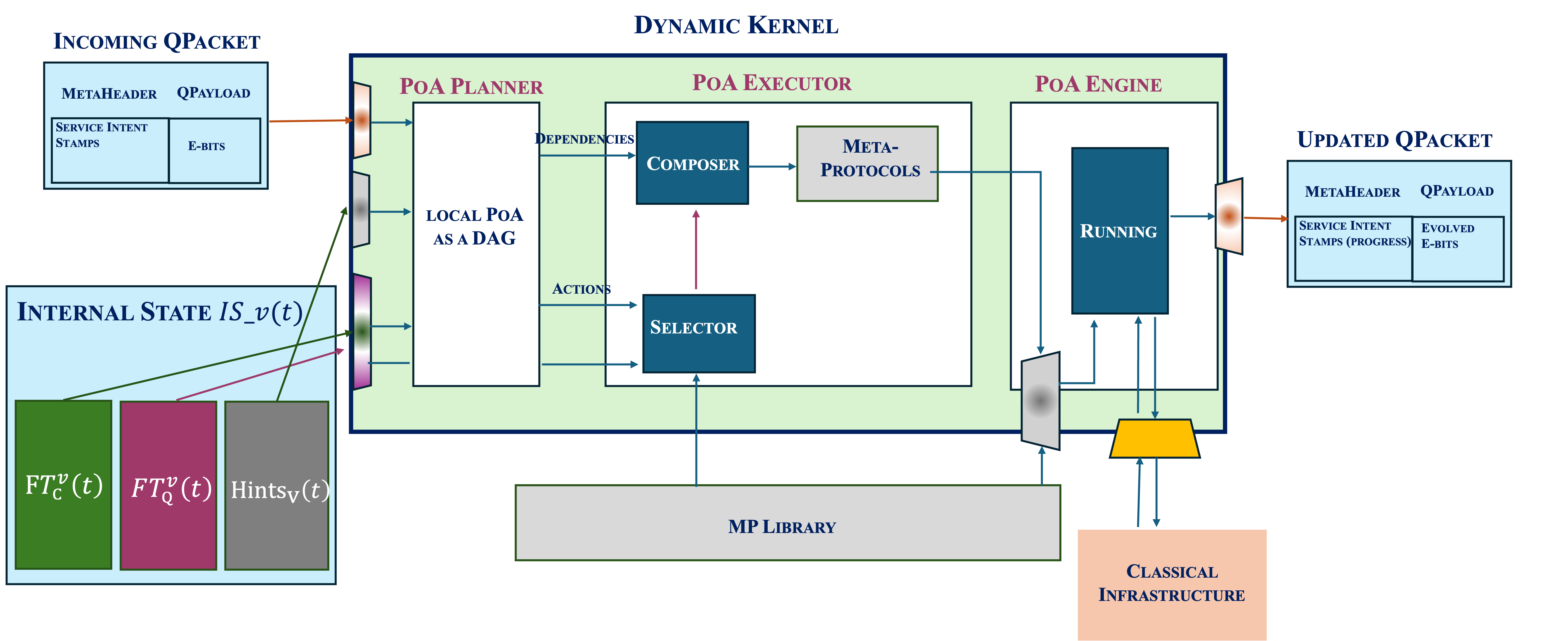}
    \caption{The dynamic kernel pipeline for quantum packet processing. Each node receives a packet carrying the service intent and accumulated stamps. The kernel computes a local Plan of Actions (PoA), selects and composes micro-protocols into meta-protocols with embedded feasibility checks, and invokes them physically. The packet is updated with new stamps and an evolved payload before being forwarded, so that the service request is gradually fulfilled hop-by-hop.}
    \label{fig:03}
    \hrulefill
\end{figure*}

\subsection{Dynamic Kernel Pipeline: Overview}
\label{sec:3.2}

Figure~\ref{fig:03} illustrates the quantum packet-processing pipeline centered on the \textit{Dynamic Kernel}, highlighting the packet structure, the three internal building blocks of the kernel, and the output packet after execution.  

\vspace{3pt}
\subsubsection{Input Quantum Packet and Local Context}
Each node processes an \textit{incoming quantum packet} composed of a \textit{quantum meta-header} and a quantum payload. The meta-header carries: (i) the \textit{service intent} introduced in eq.~\eqref{eq:02} and (ii) a sequence of \textit{stamps} that certify actions already committed along the ``service-path'' introduced in eq.~\eqref{eq:03}. The meta-header is encoded in a subspace orthogonal to the payload, enabling non-disruptive and non-perturbing readout that leaves the payload unchanged\footnote{\label{foot:04}Here, ``orthogonal subspace'' is meant in the formal quantum-mechanical sense, thereby enabling non-perturbing readout with respect to the payload state. Importantly, this does not preclude measurements on the meta-header. Rather, measurements are performed on the control subspace as required by the adopted meta-header physical implementation. By construction, such measurements act only on the meta-header and therefore do not collapse or otherwise perturb the entanglement state encoded in the payload subspace. This operational decoupling is conceptually aligned with approaches exploiting orthogonal subspaces in order to separate control and quantum-state evolution, such as for example in \cite{SimCalIll-23}.}.
The quantum payload contains entanglement qubits (e-bits) that may be consumed, transformed, or left untouched, depending on the local PoA. Payload e-bits carried in the packet do not need to be strictly tied to the current service intent.\\
Alongside the packet, the kernel consults the node \textit{internal state} $\mathcal{IS}_v(t)$, as introduced in eq.~\eqref{eq:01}. The internal state exposes via $FT_q^v(t)$ the entanglement resources incident at $v$  along with their descriptors -- such as ownership, fidelity, coherence budget, identifiers -- thereby enabling local feasibility checks and MP selection, without any reliance on global network knowledge. 

\vspace{3pt}
\subsubsection{Dynamic Kernel Components}
The Dynamic Kernel comprises three logical components: the \textit{PoA Planner}, the \textit{PoA Executor}, and the \textit{PoA Engine}. 
\begin{itemize}
    \item{\textit{PoA Planner.}}  
    The Planner constructs a local PoA from the original service intent, the stamps accumulated in the meta-header, and the node internal state $\mathcal{IS}_v(t)$. This PoA is synthesized as a local DAG $D_v(t) = \big(A_v(t),\,E_v(t)\big)$, where the vertices $A_v(t)$ are the candidate \textit{actions} the node can attempt, and the directed edges $E_v$ encode \textit{dependencies} among the actions that must be respected locally: (i) \textit{logical} causal relations (an action output feeds another); (ii) \textit{resource/quality} constraints (e.g., fidelity/coherence thresholds); and (iii) conservative, \textit{view-induced} sequencing. The resulting PoA is \textit{speculative} and \textit{local}: it captures only what node $v$ can execute to advance the service, given its own view. Importantly, the packet–kernel interface exposes the \textit{meta-header} fields explicitly. Notably, no global ``frontier'' of pending actions is carried in the packet. Instead, each node constructs its own PoA locally. This keeps the meta-header minimal and naturally adapts to dynamically evolving entanglement availability. Finally, the Planner inserts a \textit{terminal} action $a_{\mathrm{term}}\!\in\!\{\texttt{ACT\_DELIVER},\,\texttt{ACT\_FORWARD}(\cdot),\,\texttt{ACT\_DROP}\}$ to close the local DAG and eventually a soft-state action (see Sec.~\ref{sec:4.2} for further details).
    \item{\textit{PoA Executor.}} Triggered by the PoA produced by the Planner, the Executor maps \textit{candidate actions} to concrete \textit{MP/MeP} realizations, by selecting the appropriate MPs and by producing a runnable schedule. To this end, the Executor considers both (i) the actions themselves, driving the \textit{selection step}, and (ii) their dependencies, guiding the \textit{composition step} of MPs into MePs. In this way, sequencing constraints are naturally enforced prior to invocation. The Executor also binds concrete parameters -- such as timeouts or retry/backoff budgets -- based also on the node internal state, and schedules the resulting MePs under local timing/coherence constraints. Unlike the Planner, the Executor has visibility over the set of MPs actually available at node $v$ and therefore performs additional feasibility/capability checks, such as policy-enforcement constraints. Embedding these checks within the Executor ensures that only executable and authorized actions are dispatched, thereby preventing wasted resources and preserving consistency across nodes. In essence, the MPs composition and ordering adapt to both local resource state and policy.
    \item{\textit{PoA Engine.}} The Engine actually runs the scheduled MePs/MPs identified by the Executor, by invoking the underlying communication and operation primitives. The Engine enforces \textit{action boundaries}: upon a terminal outcome (commit or explicit abort), it atomically updates the \textit{meta-header}, by appending the corresponding stamp, and it updates when needed the \textit{quantum payload} and $FT_q(v)$ to reflect the entanglement processing prescribed by the PoA. No speculative writes occur: only committed outcomes are recorded, thereby ensuring monotonicity of the stamp log and consistency between control and data evolution. This separation of concerns mirrors the compiler/runtime split in classical computational systems: the Planner plans, the Executor orchestrates, by mapping and scheduling, while the Engine runs and certifies.
\end{itemize}

\vspace{3pt}
\subsubsection{Side Interfaces and Output Quantum Packet}
The kernel relies on two auxiliary interfaces: 
\begin{itemize}
    \item{\textit{MP Library Interface}:} It exposes the local catalog of available MPs. The Executor queries the library to bind actions to concrete MP/MeP realizations. 
    \item{\textit{Classical side-channel interface}:} It provides reliable classical signaling toward network nodes. Indeed, adopting SDN terminology, this interface plays both southbound and northbound roles, enabling bidirectional information flow. In particular, EDC hints may also be conveyed via this classical interface and used to refine quantum packet processing. This is not explicitly depicted in Fig.~\ref{fig:03}.
\end{itemize}  

\vspace{3pt}
After the execution described above, the kernel emits an \textit{updated packet}: the meta-header retains the original service intent, but extends the set of new stamps to certify freshly committed actions. The quantum payload is evolved accordingly and the updated packet is then eventually forwarded to the next hop selected by the local PoA, where the process repeats.

\begin{remark}
    This modular pipeline is designed for scalability and correctness, in the inherently stateful, non-local regime of entanglement. By carrying only the service intent and the monotone (append-only) sequence of stamps, rather than a speculative frontier of pending work, the packet remains compact and avoids redundancy. Each node constructs its PoA locally, adapting naturally to entanglement availability and coherence-time budgets. Embedding feasibility/capability checks in the Planner and Executor enforces safety and ordering before the actual MPs invocation: the Planner encodes logical and resource guards in the PoA, and the Executor performs capability/feasibility checks when binding actions to MPs/MePs. By isolating the PoA Engine, hardware-specific details and execution are abstracted. In this way, the kernel unifies packet-carried control with node-local autonomy, providing a quantum-native alternative to classical layering. 
\end{remark}

\subsection{Dynamic Kernel Pipeline: Formal Design}
\label{sec:3.3}

According to the above, we formalize the design of the \textit{Dynamic Kernel} as follows.\\
\textbf{Designed Dynamic Kernel:} At node $v$ and (logical) time $t$, the \textit{dynamic kernel} $\mathcal{K}_v$ is the local \textit{orchestration and execution environment} that advances a service intent by planning, composing, and invoking micro-protocols accordingly to the internal state of the node and to the in-band meta-header carried within the quantum packet.  $\mathcal{K}_v$ is not a single processing unit, but a structured tuple of logical components and interfaces:
\begin{align}
    \label{eq:04}
    \mathcal{K}_v =  \big\langle &
        \textsf{Planner}_v,\ \textsf{Executor}_v,\ \textsf{Engine}_v,\ \mathcal{I}^{\mathrm{pkt}}_v,\ \mathcal{I}^{\mathrm{state}}_v,\ \\&\nonumber
        \mathcal{I}^{\mathrm{mp}}_v,\ \mathcal{I}^{\mathrm{sig}}_v \big\rangle.
\end{align}
Specifically, the kernel exposes four first-class interfaces:
\begin{itemize}
    \item $\mathcal{I}^{\mathrm{pkt}}_v$: packet-access-interface. Reading of the meta-header $H=(\mathcal{I},\textsf{Stamps})$ occurs in a subspace orthogonal to the payload (non-perturbing).
    \item $\mathcal{I}^{\mathrm{state}}_v$: access interface to the internal state $\mathcal{IS}_v(t)=(FT_c^v(t),FT_q^v(t),\mathrm{Hints}_v(t))$, i.e., classical/quantum forwarding tables and optional EDC hints;
    \item $\mathcal{I}^{\mathrm{mp}}_v$: access interface to the locally available MP library for composing MPs into MePs.
    \item $\mathcal{I}^{\mathrm{sig}}_v$: access interface to reliable classical side-channel used by specific MPs/MePs as needed.
\end{itemize}
Formally, $\mathcal{K}_v$ realizes the following mapping:\\
\noindent\textbf{Kernel mapping.} Given an input triple $(H,\ \text{payload},\ \mathcal{IS}_v(t))$, where
$H=(\mathcal{I},\textsf{Stamps})$ is the meta-header, \textit{payload} is the carried quantum state(s),
and $\mathcal{IS}_v(t)$ is the node internal state, the kernel realizes the mapping:
\begin{equation}
\label{eq:05}
\mathcal{K}_v:\ \big(H,\ \text{payload},\ \mathcal{IS}_v(t)\big)\ \longmapsto\
\big(H',\ \text{payload}',\mathcal{IS}'_v(t)\big).
\end{equation}

\medskip
\noindent\textbf{Structure and operation.} The logical components in \eqref{eq:04} are executed in the following order:
\begin{enumerate}
    \item \textit{PoA Planner} ($\textsf{Planner}_v$) builds a speculative local \textit{Plan of Actions} as a DAG $D_v(t)=(A_v(t),E_v(t))$, given $(\mathcal{I},\textsf{Stamps},\mathcal{IS}_v(t))$, accessed via $\mathcal{I}^{\mathrm{pkt}}_v$ and $\mathcal{I}^{\mathrm{state}}_v$.
    \item \textit{PoA Executor} ($\textsf{Executor}_v$): performs feasibility/capability checks using $\mathcal{IS}_v(t)$ and $\mathcal{I}^{\mathrm{mp}}_v$, it selects a feasible subset of actions and maps each action to a MP/MeP (composition of MPs) realization. It produces a schedule that honors dependencies $E_v(t)$ and policy constraints.
    \item \textit{PoA Engine} ($\textsf{Engine}_v$): invokes the scheduled MPs (eventually using $\mathcal{I}^{\mathrm{sig}}_v$ when classical signaling is required), updates the quantum payload and $FT_q(v)$ as prescribed, and appends a new stamp to $H$ upon each \textit{action commit}.
\end{enumerate}

\medskip
\noindent\textbf{Guarantees.} (i) \textit{Orthogonal read}: $H$ is read in a subspace orthogonal to the payload, thereby enabling non-perturbing access. (ii) \textit{Monotonicity}: only \textit{committed} actions are appended as stamps (append-only log). (iii) \textit{Commit-Atomicity}: header/payload/$FT_q(t)$ updates become visible atomically at action commit. On abort, no update is visible. (iv) \textit{Locality}: no global state is assumed. Indeed, each hop recomputes its PoA from $(\mathcal{I},\textsf{Stamps},\mathcal{IS}_v(t))$. (v) \textit{Minimality}: $H'$ carries only the service intent and stamps, providing the next hop the minimal and consistent context to compute its own PoA. (vi) \textit{Single-writer} for the meta-header: At any time, exactly one node holds the authoritative meta-header $H$. Only the holder may append stamps at action commit. Authority can be transferred by forwarding $H$ to the next node that will coordinate the next commit\footnote{When EDC hints enable parallel operations at different nodes, for example two non-conflicting SWAPs, the single-writer discipline is preserved by policy. Although detailed coordination mechanisms are beyond the scope of this paper, two standard solutions could be: (i) the current holder of $H$ supervises the parallel actions via classical-side channel and appends one stamp per completion upon receiving acknowledgments; (ii) $H$ is forwarded to each action coordinator in turn, which appends its stamp upon completion. In both cases, execution proceeds in parallel while stamping remains single-writer and the global commit DAG stays acyclic.}.

The algorithmic description of the kernel pipeline on a per-node basis is given in Algorithm~\ref{alg:01}.

\begin{remark}
    In the proposed architecture, action feasibility is assessed locally through the described kernel pipeline. As a result, many potential resource discrepancies are filtered out prior to invocation. When they cannot be ruled out a-priori, they are resolved through the operational outcome of the invoked MPs/MePs, without requiring any global synchronization. If, at execution time, a required quantum resource is stale, no longer available or no longer satisfies fidelity/coherence constraints, the corresponding action simply fails locally and is handled via the retry/backoff/abort logic captured in Algorithm~1. In other words, the \textit{Dynamic Kernel} operating at a single node does not attempt to maintain or reconcile a global view of quantum resources. When an action involves multiple nodes (e.g., entanglement generation, purification, or swapping), correctness is ensured by the commit discipline: an action commits -- and thus becomes visible to subsequent nodes -- only when all required conditions have been satisfied according to the semantics of the enclosing MP/MeP. Hence, in the proposed architecture, distributed MPs may indeed succeed at some nodes and fail at others. However, such outcomes remain confined within the enclosing MP/MeP and never become externally visible. An action is certified and stamped only if the corresponding MeP reaches a commit according to its semantics. Partial success or asymmetric MP outcomes without a corresponding commit simply result in local failure handling and do not induce any global inconsistency. Importantly, discrepancies in quantum resources -- such as partial consumption of entanglement or asymmetric success of distributed MPs -- are operationally indistinguishable from decoherence effects or loss. They are therefore handled uniformly as a local feasibility failure within the proposal, rather than as global inconsistencies requiring reconciliation.\\
    The same ``containment'' principle applies to partial failures of auxiliary classical signaling. If the required classical information does not arrive within policy-defined bounds, the enclosing MP/MeP cannot satisfy its commit conditions and the action does not commit. The failure is handled locally at the node without exporting inconsistent state via stamps.
\end{remark}

From the above discussion, it also emerges that the meta-header growth is \textit{architecturally bounded} and \textit{semantically constrained}: it scales only with the number of \textit{committed actions} (stamps), and not with retries, failed attempts, speculative steps, or transient protocol states. Since stamps are appended exclusively at \textit{action-commit boundaries}, the meta-header size reflects the logical progress of a service, rather than its operational complexity or runtime variability. This yields a predictable and bounded overhead, independent of network size. Consequently, for any given service intent, the number of committed actions is \textit{finite and predictable}, as it corresponds to the minimal set of certified milestones required to reach a terminal outcome. In other words, in the proposed suite, a \textit{service intent} is a semantically well-defined objective with commit-based termination, and the finiteness of the stamp sequence follows directly from the commit discipline underlying service fulfillment. Additional mechanisms, as instance stamp compaction, could be introduced as optimization layer on top of the proposed design without altering its semantics. Such mechanisms are natural future directions rather than prerequisites for scalability.

\begin{algorithm}[t]
\caption{Dynamic Kernel at node $v$}
\label{alg:01}
\DontPrintSemicolon

\KwIn{\parbox[t]{0.6\linewidth}{Quantum Packet $p=(H,\text{payload})$ with $H=(\mathcal{I},\textsf{Stamps})$ via $\mathcal{I}^{\mathrm{pkt}}_v$; internal state $\mathcal{IS}_v(t)$ via $\mathcal{I}^{\mathrm{state}}_v$.}}
\BlankLine
\KwOut{Updated packet $p'=(H',\text{payload}')$ and internal state.}

\BlankLine
\textbf{1) Orthogonal read (packet I/O):} 
$\mathcal{I} \gets \mathcal{I}^{\mathrm{pkt}}_v.\textsc{ReadIntent}()$; 
$\textsf{Stamps} \gets \mathcal{I}^{\mathrm{pkt}}_v.\textsc{ReadStamps}()$ \Comment*{non-perturbing}

\textbf{2) PoA Planner (MP Agnostic):} build a local action DAG $D_v=(A_v,E_v)$ from $(\mathcal{I},\textsf{Stamps},\mathcal{IS}_v)$.\;
$D_v=(A_v,E_v)\gets \textsc{PlanPoA}\big(\mathcal{I},\textsf{Stamps},\mathcal{I}^{\mathrm{state}}_v\big)$\;
Insert a terminal action in $A_v$ and eventually soft-state action(s): $a_{\mathrm{term}}\in\{\texttt{ACT\_DELIVER},\,\texttt{ACT\_FORWARD}(\widehat{nh}),\,\texttt{ACT\_DROP}\}$ (here $\widehat{nh}$ is a placeholder).

\textbf{3) PoA Executor (feasibility, mapping, schedule):}\;
$S_v \gets \textsc{SelectFeasible}\big(D_v,\mathcal{I}^{\mathrm{state}}_v\big)$ \Comment*[r]{select planner-produced actions executable now, \textit{respecting} $E_v$ and checking capabilities/coherence/policy}
$\Phi \gets \textsc{MapAndBind}\big(S_v,\mathcal{I}^{\mathrm{mp}}_v,\mathcal{I}^{\mathrm{state}}_v\big)$ \Comment*[r]{for each action $a\in S_v$, map $a$ to a MeP (sequence of MPs) and bind concrete params (e.g., $\widehat{nh}\mapsto nh$)}
$\Sigma \gets \textsc{Schedule}\big(\Phi,E_v,\mathcal{I}^{\mathrm{state}}_v\big)$ \Comment*[r]{compute an execution order honoring $E_v$ and resource/timing constraints}

\textbf{4) PoA Engine (invoke MPs; commit on action boundaries):}\;
\ForEach{$(a,\mathrm{MeP}_a)\in\Sigma$}{
  \ForEach{$\mathrm{MP}\in \mathrm{MeP}_a$}{
    \eIf{$\mathcal{I}^{\mathrm{mp}}_v.\textsc{InvokeMP}(\mathrm{MP},\mathcal{I}^{\mathrm{sig}}_v)$}{
      $\mathcal{I}^{\mathrm{state}}_v.\textsc{UpdateFTqAndPayload}()$\;
    }{
      \textsc{HandleFailure}(retry / backoff / abort)\;
      \If{\textsc{Abort}}{
        $\mathcal{I}^{\mathrm{pkt}}_v.\textsc{AppendStamp}(\textsc{FailureStamp}(a))$;\ \KwRet drop $p'$\;
      }
    }
  }
  $\mathcal{I}^{\mathrm{pkt}}_v.\textsc{AppendStamp}(\textsc{CommitStamp}(a))$ \Comment*[r]{atomic action commit}
  \If{$a=\texttt{ACT\_DELIVER}$}{\KwRet deliver $p'$}
  \If{$a=\texttt{ACT\_FORWARD}$ with $nh$}{\KwRet forward $p'$ to $nh$}
  \If{$a=\texttt{ACT\_DROP}$}{\KwRet drop $p'$}
}
\end{algorithm}

\subsection{Fundamental micro-protocols (MPs)}
\label{sec:3.4}

The kernel operates on a library of \textit{MPs} that capture
elementary, reusable primitives, on which MePs are composed. We identify a \textit{minimal} and \textit{illustrative} set of MPs, without claiming a universal catalog. Indeed, as technology advances, procedures that are today realized as MePs may collapse into single MPs. \textit{The suite is deliberately invariant\footnote{The authors thank John Day for his seminal work on abstraction and invariance, which inspired this principle.} to such boundary shifts}: the \textit{Planner} is MP-agnostic, the \textit{Executor} maps actions to the currently available MP library and binds concrete parameters; the \textit{Engine} invokes the MPs, without changing the meta-header or stamps semantics. Thus, planning and execution treat MPs as opaque primitives with declared pre/post conditions and timing/coherence constraints, so correctness and scalability do not depend on where the MP/MeP cut is drawn.
\begin{enumerate}
    \item \textbf{MP\textsubscript{GEN}} (single entanglement-generation attempt): performs one attempt to create link-level entanglement on a physical quantum interface, either heralded or un-heralded.
    \item \textbf{MP\textsubscript{QP}} (quantum processing): executes local quantum operations, such as single- and multiple-qubit gates, as well as related primitives.
    \item \textbf{MP\textsubscript{SYN}} (synchronization): maintains/refreshes time/phase references required by, for example, entanglement generation process, swapping, purification, or entanglement-consumption MePs.
    \item \textbf{MP\textsubscript{SIG}} (classical signaling): delivers minimal classical side information, such as Bell State measurements, ACK/NACK messages, policy acknowledgments, with reliability, ordering, and latency properties consistent with the node’s internal state and the service intent.
    \item \textbf{MP\textsubscript{FW}} (packet forwarding): advances fulfillment of the service intent by delivering the meta-header and, when beneficial, the payload (including payload not tied to the current intent) to the selected next hop $nh$.
\end{enumerate}
\noindent\textit{Extensibility.} Additional MPs can be introduced over time without changing the kernel’s design: only the MP library and the Executor’s mapping logic require updates. The meta-header and the end-to-end service-fulfillment semantics remain unchanged.

\subsection{Node Roles and End-to-End Perspective}
\label{sec:3.5}

The pipeline operates uniformly at every node, but its behavior depends on the current stage of service fulfillment.

\begin{itemize}
    \item Initiating Node: The originator constructs the initial PoA, by using only its internal state. Since no actions have yet been committed, the stamps field of the meta-header is empty. Accordingly, the packet carries only the service intent and (optionally) an initial payload. By design, no frontier field of pending steps is included, thereby avoiding redundancy and any reliance on global knowledge. 
	\item Intermediate Nodes: When the packet reaches an intermediate node selected by the previous hop PoA, the service intent is not yet fully satisfied. The node consults its internal state and it recomputes its own PoA from $(\mathcal{I},\textsf{Stamps},\mathcal{IS}_v(t))$. It selects a feasible subset of actions, executes them, appends the resulting stamps to the meta-header, and evolves the payload/$FT_q^v(t)$ accordingly, without requiring global network knowledge. Each node contributes only its portion of the progress it can safely realize given its internal state, current availability, and connectivity. This service-oriented operation dispenses with global synchronization and is naturally aligned with the stateful nature of entanglement resources.
    \item Terminating node: When the service intent is satisfied at $v$, the kernel commits a final stamp certifying completion and delivers locally (thus, no further forwarding is performed).
\end{itemize}
\noindent \textit{End-to-end perspective.} Under this model, the service request is fulfilled as a \textit{dynamic global DAG} that emerges from local commits (local refinements) at each node rather than being precomputed or explicitly carried. Each hop updated the packet with new stamps that certify completed actions, while the payload eventually evolves in lockstep. By recording only completed steps rather than speculative futures, the meta-header remains minimal, ensures correctness, and avoids the scalability limits associated with distributing a fully precomputed global plan. When EDC hints are available, independent actions at different nodes may proceed in parallel; in their absence, conservative, view-induced ordering preserves safety without requiring global state.
\begin{remark}
    One might argue that, since stamps encode classical information, they could be carried entirely over a parallel classical network infrastructure. We observe that this would reintroduce the very limitations of a layered classical architecture. Out-of-band signaling decouples the control from the evolving quantum state, forcing global coordination to maintain consistency. The system would either require global synchronization or risk divergence between the quantum substrate and its classical description. By contrast, embedding the meta-header in-band with the quantum packet preserves consistency between state evolution and control, ensures that local refinements remain coherent, and avoids the scalability pitfalls of a purely classical control-plane.
\end{remark}

\begin{remark}
    The proposed architecture does not assume arbitrarily long coherence times, nor does it rely on planning/execution latencies being negligible. As in any quantum-networking design, a minimal physical prerequisite is the availability of quantum memories whose coherence time suffices to complete the intended local operations; without such capability, no network-level protocol (layered or otherwise) can function. Within these unavoidable constraints, coherence budgets are treated as first-class feasibility parameters: they are consulted during PoA construction and enforced during execution. If an MP/MeP fails due to decoherence or timing violations, the enclosing action simply does not reach commit and therefore produces no stamp; recovery is handled locally via retry, backoff, or abort according to policy, without compromising global correctness. In this sense, the suite remains robust to timing variability by certifying only completed actions, rather than presuming success.
\end{remark}

\section{Service Fulfillment as an Emergent Commit DAG}
\label{sec:04}

This section formalizes end-to-end service progression as a direct \textit{consequence} of the design principles introduced in Secs.~\ref{sec:02}–\ref{sec:03}: local autonomy, in-band coordination via the meta-header $H=(\mathcal{I},\textsf{Stamps})$, and per-node composition by the Dynamic Kernel $\mathcal{K}_v$ (Eq.~\eqref{eq:03}, Alg.~\ref{alg:01}). We discuss how the monotone accumulation of stamps induces a network-wide partial order over completed actions and, therefore, an emergent DAG certifying service fulfillment, without distributing a global plan or requiring any network-wide synchronization.

As described in Sec.~\ref{sec:03}, within the kernel an \textit{action} is planned by the \textsf{Planner}, mapped to a MeP by the \textsf{Executor}, and physically realized by the \textsf{Engine}. The completion of an action -- the action commit - occurs when the Engine reaches a terminal outcome (success or explicit failure) and appends a
\textit{stamp} to the meta-header. Clearly, the action execution may involve multiple MPs, retries, or distributed coordination, but only the commit event is recorded. Commits are irreversible and stamps accumulate monotonically along the service path, thereby certifying only terminal outcomes.

Thus, by design, a \textit{stamp} is the in-band certification that an action has committed. A stamp can have a minimal structure as follows:
\begin{equation}
    \label{eq:06}
    s = (a,\,\mathrm{Supp}(a),\,T(a),\,\mathrm{EntIDs},\,\text{outcome},\,\mathrm{\xi}),
\end{equation}
where $a$ is the action identifier, $\mathrm{Supp}(a)$ is the set of participating nodes, $T(a)$ denotes the commit timestamp (physical or logical), $\mathrm{EntIDs}$ denotes the entanglement resources eventually consumed/produced, \text{outcome} represents the success or failure of the action execution, and $\xi$ aggregates optional metadata.

Accordingly, the end-to-end fulfillment of a service intent emerges as \textit{the distributed construction of a dynamic global DAG of action commits}. This abstraction decouples the logical progression of service fulfillment from the physical implementation of actions, whose MeP/MP implementations may include internal loops, retries, and inter–node coordination\footnote{For example, if the required EPR pairs are already available with sufficient quality, entanglement purification can be realized by a single MP, i.e., \textbf{MP\textsubscript{QP}}. Otherwise it expands into a MeP that iterates \textbf{MP\textsubscript{SYN}} and \textbf{MP\textsubscript{GEN}} with \textbf{MP\textsubscript{QP}}, interleaving \textbf{MP\textsubscript{SIG}} as needed, until the target quality is reached. Regardless from the particular case, the packet carries only minimal history (intent + stamps), while service fulfillment -- captured by the growing sequence of stamps -- advances hop-by-hop.}.

Edges in the global DAG capture only the \textit{necessary happens-before constraints} (namely, the causal constraints) between committed actions, as determined by logical, resource, or conservative orderings induced by local views. These dependencies are not injected by any single node, but they emerge systemically from the distributed evolution under local node-views. Indeed, in the absence of EDC support, the distributed evolution (by design) of the service intent may be forced into conservative orderings: each node, relying only on its local view, assumes dependencies that may not exist globally. This manifests as additional ``view-induced edges'' that serialize otherwise independent actions, as instance two SWAP operations on disjoint entanglement resources. With EDC hints, some of these conservative edges can be pruned, and the DAG reveals the true concurrency inherent in the service. Thus, the global DAG is best understood as a partial order over action commits, whose degree of concurrency depends on the quality and the scope of control-plane information available to nodes. This is formalized in the following subsection.

\subsection{Global DAG}
\label{sec:04.1}

Let $\mathsf{S}(t)$ be the set of all the \textit{stamps} that have been appended to the meta-header across the network nodes up to (logical) time $t$. We express the emergent global graph as follows.

\textbf{Emerging Global DAG}
The emergent global graph at time $t$ is a direct acyclic graph (DAG) defining the dynamic global order of executed actions fulfilling the service intent:
\begin{equation}
    \label{eq:07}
    D^*(t) \;=\; \big(A^*(t),\,E^*(t)\big),
\end{equation}
with vertices indicating the actions actually committed
(by producing a stamp) by time $t$:
\begin{align}
    \label{eq:06}
    A^*(t) = \{\alpha: & \alpha \text{ has committed by time } t \nonumber \\
    & \text{ and produced a stamp}\},
\end{align}
and edges capturing only \emph{commit-to-commit} dependencies:
\begin{align}
    \label{eq:07}
(\alpha,\beta) & \in E^*(t) \nonumber \\
    & \iff \beta \ \text{cannot commit unless }\alpha\ \text{has committed}.
\end{align}

It is worth stressing that the acyclic nature of $D^*(t)$ is by construction: edges of $E^*(t)$ point from earlier to later commits and cycles cannot form as a consequence of design, which is characterized by the following features:
\begin{enumerate}
    \item[\textup{(i)}] \textit{Commit-only stamping:} a stamp is emitted and appended to the meta-header \textit{only} when the PoA Engine completes (commits) an action instance. An abort emits a terminal failure stamp.
    \item[\textup{(ii)}] \textit{Monotone timestamps:} each stamp carries a timestamp $T(\cdot)$ that is monotone with respect to the local happens-before relation ($a \rightarrow b \Rightarrow T(a) < T(b)$, as instance via Lamport or vector clocks.
    \item[\textup{(iii)}] \textit{Irreversibility:} commits are permanent (no un-commit).
    \item[\textup{(iv)}] \textit{Edge minimality:} $E^*(t)$ contains only dependencies required for commit. Local, speculative planning edges $E_v(t)$ produced by PoA Planners are not exported in $E^*(t)$.
\end{enumerate}
Thus, $D^*(t)$ is a DAG and induces a partial order over committed actions up to time $t$. In this sense, $D^*(t)$ is an \textit{emergent} global structure induced solely by packet-carried stamps. 

\begin{remark}[Engine “zoom-in”]
    It is worthwhile to observe that each vertex of $D^*(t)$ abstracts the \textit{implementation graph} executed by the local PoA Engine to realize a local node action: a composition of MPs that may include loops -- due to retries and/or backoffs -- and inter-node coordination. This per-action graph need not be acyclic. The commit DAG $D^*(t)$ deliberately \emph{decouples} such intra-action control-flow: only terminal commits materialize as vertices, ensuring that $D^*(t)$ itself remains a DAG. Zooming into a vertex in $D^*(t)$ reveals this finer-grained intra-action control-flow.
\end{remark}

\noindent\textit{Concurrency and the Role of EDC.} In the absence of EDC hints, nodes may enforce conservative (view-induced) orderings that serialize otherwise independent actions, thereby reducing visible parallelism. With EDC guidance, those spurious constraints are pruned, and $D^*(t)$ exposes the true parallelism.

\subsection{Case Study: End-to-End Teleportation $A\!\rightarrow\!B$}
\label{sec:4.2}

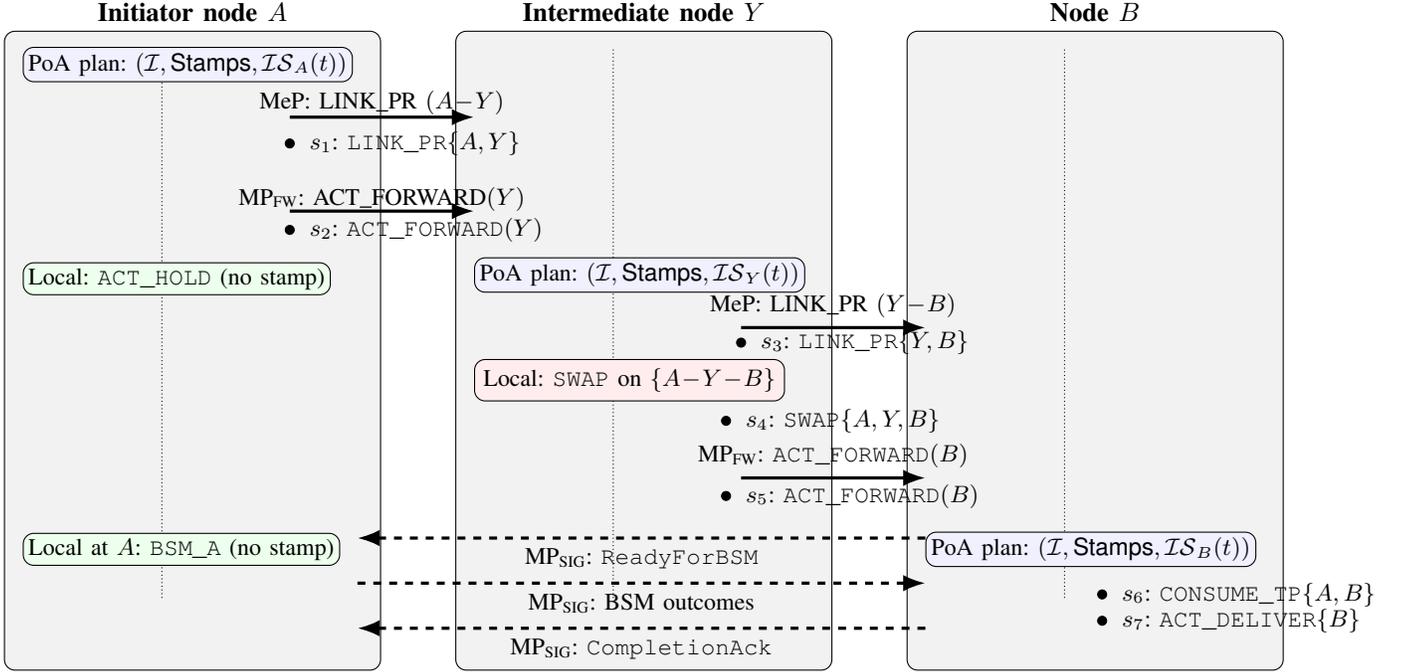
\begin{figure*}[t]
\centering
\begin{tikzpicture}[
  font=\small,
  >=Latex,
 lane/.style={draw, rounded corners, fill=gray!10, minimum width=5.0cm, minimum height=8.5cm},
  title/.style={font=\bfseries},
  event/.style={draw, rounded corners, fill=blue!6, inner sep=2pt},
  local/.style={draw, rounded corners, fill=green!7, inner sep=2pt},
  msg/.style={-Latex, very thick},
  classical/.style={-Latex, dashed, very thick},
  stampdot/.style={circle, draw, fill=black, inner sep=1.2pt},
  labelr/.style={anchor=west, align=left},
  labell/.style={anchor=east, align=right}
]

\def\xA{0}
\def\xY{6.0}
\def\xB{12.0}

\def\ytop{0.2}
\def\ybot{-7.6}

\node[lane, anchor=north west] (laneA) at (\xA,\ytop) {};
\node[lane, anchor=north west] (laneY) at (\xY,\ytop) {};
\node[lane, anchor=north west] (laneB) at (\xB,\ytop) {};

\node[title, anchor=north, yshift=5mm] at ($(laneA.north west)!0.5!(laneA.north east)$) {Initiator node $A$};
\node[title, anchor=north, yshift=5mm] at ($(laneY.north west)!0.5!(laneY.north east)$) {Intermediate node $Y$};
\node[title, anchor=north, yshift=5mm] at ($(laneB.north west)!0.5!(laneB.north east)$) {Node $B$};

\draw[densely dotted] (\xA+2.1,\ytop-0.25) -- (\xA+2.1,\ybot+0.25);
\draw[densely dotted] (\xY+2.1,\ytop-0.25) -- (\xY+2.1,\ybot+0.25);
\draw[densely dotted] (\xB+2.1,\ytop-0.25) -- (\xB+2.1,\ybot+0.25);

\node[event, anchor=west] (Aplan) at (\xA+0.25,-0.25) {PoA plan: $(\mathcal{I},\textsf{Stamps},\mathcal{IS}_A(t))$};

\draw[msg] (\xA+3.8,-0.95) -- node[above,sloped,yshift=-1mm]{MeP: LINK\_PR $(A\!-\!Y)$} (\xY+0.25,-0.95);

\node[stampdot] (s1) at (\xA+3.8,-1.3) {};
\node[labelr] at ($(s1)+(0.14,0)$) {$s_1$: \texttt{LINK\_PR}\{$A,Y$\}};

\draw[msg] (\xA+3.8,-2.2) -- node[above,sloped,yshift=-1.3mm]{MP\textsubscript{FW}: ACT\_FORWARD$(Y)$} (\xY+0.25,-2.2);
\node[stampdot] (s2) at (\xA+3.8,-2.45) {};
\node[labelr] at ($(s2)+(0.14,0)$) {$s_2$: \texttt{ACT\_FORWARD}$(Y)$};

\node[local, anchor=west] (Ahold) at (\xA+0.25,-3.1) {Local: \texttt{ACT\_HOLD} (no stamp)};

\node[event, anchor=west] (Yplan) at (\xY+0.25,-3.05) {PoA plan: $(\mathcal{I},\textsf{Stamps},\mathcal{IS}_Y(t))$};

\draw[msg] (\xY+3.8,-3.75) -- node[above,sloped]{MeP: LINK\_PR $(Y\!-\!B)$} (\xB+0.25,-3.75);

\node[stampdot] (s3) at (\xY+3.8,-3.95) {};
\node[labelr] at ($(s3)+(0.14,0)$) {$s_3$: \texttt{LINK\_PR}\{$Y,B$\}};

\node[draw, rounded corners, anchor=west,fill=red!7] (Yswap) at (\xY+0.25,-4.45) {Local: \texttt{SWAP} on \{$A\!-\!Y\!-\!B$\}};
\node[stampdot] (s4) at (\xY+3.6,-4.99) {};
\node[labelr] at ($(s4)+(0.14,0)$) {$s_4$: \texttt{SWAP}\{$A,Y,B$\}};

\draw[msg] (\xY+3.8,-5.75) -- node[above,sloped]{MP\textsubscript{FW}: \texttt{ACT\_FORWARD}$(B)$} (\xB+0.25,-5.75);
\node[stampdot] (s5) at (\xY+3.6,-5.99) {};
\node[labelr] at ($(s5)+(0.14,0)$) {$s_5$: \texttt{ACT\_FORWARD}$(B)$};

\node[event, anchor=west] (Bplan) at (\xB+0.25,-6.7) {PoA plan: $(\mathcal{I},\textsf{Stamps},\mathcal{IS}_B(t))$};

\draw[classical] (\xB+0.25,-6.55) -- node[below,sloped]{MP\textsubscript{SIG}: \texttt{ReadyForBSM}} (\xA+4.7,-6.55);

\node[local, anchor=west] (Absm) at (\xA+0.25,-6.7) {Local at $A$: \texttt{BSM\_A} (no stamp)};
\draw[classical] (\xA+4.7,-7.15) -- node[below,sloped]{MP\textsubscript{SIG}: BSM outcomes} (\xB+0.25,-7.15);

\node[stampdot] (s6) at (\xB+2.6,-7.3) {};
\node[labelr] at ($(s6)+(0.14,0)$) {$s_6$: \texttt{CONSUME\_TP}\{$A,B$\}};
\node[stampdot] (s7) at (\xB+2.6,-7.65) {};
\node[labelr] at ($(s7)+(0.14,0)$) {$s_7$: \texttt{ACT\_DELIVER}\{$B$\}};

\draw[classical] (\xB+0.25,-7.75) -- node[below,sloped]{MP\textsubscript{SIG}: \texttt{CompletionAck}} (\xA+4.7,-7.75);

\end{tikzpicture}
    \caption{Swim-lane execution of the Dynamic Kernel for $\texttt{TELEPORT}(A\!\rightarrow\!B)$ across $A$–$Y$–$B$. At each hop the kernel reads intent $\mathcal{I}$ and stamps $\textsf{Stamps}$, recomputes a local PoA, composes MPs/MePs, and invokes MPs. Stamps are appended only at action commits. Retries and backoffs remain local to MePs and do not inflate the meta-header. \texttt{ACT\_HOLD} and \texttt{BSM\_A} are local and produce no stamps. Meta-header authority transfers on \texttt{ACT\_FORWARD}. Dashed arrows are classical signaling.}
    \label{fig:04}
    \hrulefill
\end{figure*}

\begin{table}[!t]
    \centering
    \renewcommand{\arraystretch}{1.5}
    \caption{Excerpt of stamps for \texttt{TELEPORT} $A\!\rightarrow\!B$ (action-level commits) illustrated in Fig.~\ref{fig:04}.}
    \begin{tabular}{l l l l}
        \hline
        \hline
        ID & Action & Support & Outcome \\
        \hline
        $s_1$ & \texttt{LINK\_PREP}$(A\!-\!Y)$ & $\{A,Y\}$ & \textsf{ok} \\
        $s_3$ & \texttt{LINK\_PREP}$(Y\!-\!B)$ & $\{Y,B\}$ & \textsf{ok} \\
        $s_4$ & \texttt{SWAP}$(A\!-\!Y\!-\!B)$ & $\{A,Y,B\}$ & \textsf{ok} \\
        $s_6$ & \texttt{CONSUME\_TP}$(A\!\rightarrow\!B)$ & $\{A,B\}$ & \textsf{ok} \\
        $s_7$ & \texttt{ACT\_DELIVER} & $\{B\}$ & \textsf{ok} \\
        \hline
        \hline
    \end{tabular}
    \label{tab:02}
\end{table}

This section shows how the suite advances a service intent using only local state and in-band coordination. At each hop, the dynamic kernel recomputes a local PoA from the node internal state and the packet meta-header; stamps are appended only at action-commit boundaries; and the end-to-end execution order emerges as a global service-fulfillment DAG, without distributing any precomputed global plan. As a case study, we instantiate the intent $\texttt{TELEPORT}(A\!\rightarrow\!B)$ over a two-hop entanglement overlay $A\!-\!Y\!-\!B$. Table~\ref{tab:02} lists an excerpt of stamps, while \textit{Fig.~\ref{fig:04}} shows the kernel pipeline in a swim-lane view.

\vspace{3pt}
\subsubsection{Initiation at $A$} 
The application at $A$ issues a service intent requesting teleportation from $A$ to $B$ with target thresholds for end-to-end fidelity $F_{t}$ and residual coherence budget $\tau_{\min}$:
\begin{equation}
    \mathcal{I}=(\textsf{TELEPORT}, QA_A\!\rightarrow\!QA_B;\,F_{t},\tau_{\min}),
\end{equation}
where $QA_A$ and $QA_B$ denote the quantum addresses of the two nodes \cite{CalCac-25}.

\textit{Input packet.} The quantum packet is \(p=(H,\text{payload})\) with \(H=(\mathcal{I},\textsf{Stamps}=\langle\,\rangle)\), namely, with an empty \textsf{Stamps}. As pointed out before, the payload may carry auxiliary EPRs not necessarily used by this service.

\textit{Internal state.} Each node $v\in\{A,Y,B\}$ maintains $\mathcal{IS}_v(t)=(FT_c^v(t),FT_q^v(t),\mathrm{Hints}_v(t))$ as described in Sec.~\ref{sec:2.1}. At the initial stage, $A$ does not hold a direct $A\!\leftrightarrow\!B$ EPR\footnote{Optional EDC hints
in $\mathrm{Hints}_v(t)$ may advertise opportunities for parallel generation on disjoint interfaces at nodes $A$ and $Y$, as instance. The kernel exploits such hints without assuming a global view.}.

\textit{PoA Planner (MP-agnostic).} The Planner orthogonally reads $H$ and, by exploiting $\mathcal{IS}_A(t)$, constructs a local PoA $D_A=(A_A,E_A)$ with actions such as:
\begin{align}
    \label{eq:11}
    \{&\texttt{SYN}(A\!-\!Y), \texttt{GEN}(A\!-\!Y), 
    \texttt{PURIFY}(A\!-\!Y) \nonumber \\
    & \texttt{ACT\_FORWARD}(\widehat{Y}), \ \texttt{ACT\_HOLD}\},\end{align}
with edges $E_A$ capturing logical and quality constraints. For example, the generation-action is required multiple times when purification is needed, namely, whenever the fidelity is below the targeted threshold: $\texttt{GEN}^{\times k}\!\rightarrow\!\texttt{PURIFY}$ until $F\!\ge\!F_t$. The $\texttt{ACT\_HOLD}$ is a local, soft-state action that keeps the initiator (node $A$) waiting until an external wake-up (e.g., via classical signaling) arrives. It does not append a stamp to $H$ and it is never exported across hops. In this case-study, $\texttt{ACT\_HOLD}$ is necessary because $A$ cannot certify end-to-end completion on its own: the authoritative terminal commit is recorded at $B$ after the Pauli correction, required by the teleporting protocol. When $A$ receives $B'$s completion stamp, it releases its local soft state $\texttt{ACT\_HOLD}$. This preserves the acyclicity of the global commit DAG, since $\texttt{ACT\_HOLD}$ never materializes as a stamp.

\textit{Executor (mapping actions to MPs/MePs).} Feasibility and capabilities checks against $\mathcal{IS}_A(t)$ select the planned
actions executable now. Each selected action is then mapped to MPs/MePs composed from the MP library and bound to concrete
parameters: 
\begin{itemize}
    \item[-] \emph{A-Y Link preparation:} compose a meta-protocol $\mathrm{MeP}^{A\!-\!Y}_{\text{link}}$ as \textbf{MP\textsubscript{SYN}} $\rightarrow$ \textbf{MP\textsubscript{GEN}}$^{\times k}$ $\rightarrow$ \textbf{MP\textsubscript{QP}} (local gates and measurements used by purification), with classical feedback via \textbf{MP\textsubscript{SIG}} when required.
    \item[-] \emph{In-band forwarding:} implement \texttt{ACT\_FORWARD}($\widehat{Y}$) directly as the micro-protocol \textbf{MP\textsubscript{FW}}$(\widehat{Y}=Y)$.
\end{itemize}
As said, the \texttt{ACT\_HOLD} is installed as local soft-state. A schedule honoring $E_A$ and coherence windows is produced; retries/backoffs remain inside $\mathrm{MeP}^{A\!-\!Y}_{\text{link}}$.

\textit{Engine (invoking MPs).} The Engine interfaces with the MP library and executes the scheduled MPs in order:
\begin{itemize}
    \item[(i)] time/phase alignment via \textbf{MP\textsubscript{SYN}};
    \item[(ii)] repeated entanglement attempts via \textbf{MP\textsubscript{GEN}};
    \item[(iii)] local gates/measurements for purification via \textbf{MP\textsubscript{QP}} with any required acknowledgments via \textbf{MP\textsubscript{SIG}};
    \item[(iv)] in-band forwarding to $Y$ via \textbf{MP\textsubscript{FW}}.
\end{itemize}
Upon the \emph{commit} of each planned action, the Engine appends a stamp $s = (a,\,\mathrm{Supp}(a),\,T(a),\,\mathrm{EntIDs},\,\text{outcome},\,\mathrm{\xi})$ to $H.\textsf{Stamps}$. In this case-study, a stamp trace at $A$ could be:
\begin{align}
    s_1&=(\texttt{LINK\_PR}(A\!-\!Y),\{A,Y\},T_1,\{\mathrm{e}_{AY}\},\textsf{ok},F_{AY},\tau_{AY}), \nonumber \\
    s_2&=(\texttt{ACT\_FORWARD}(Y),\{A\},T_2,\varnothing,\textsf{ok},-). \nonumber
\end{align}
The underlying MP invocations -- with retries and backoffs -- remain encapsulated within their MePs, via Excutor policies and Engine commit boundaries. The updated packet proceeds to $Y$ carrying the refined meta-header and a purified ebit of the $A\!-\!Y$ EPR pair in the payload. Meanwhile, $A$ enters \texttt{ACT\_HOLD} locally (no stamp) awaiting a wake-up.

\vspace{3pt}
\subsubsection{Intermediate node $Y$}
The PoA Planner at node $Y$ receives $p=(H,\text{payload})$ with $H=(\mathcal{I},\textsf{Stamps})$ carrying the commits produced at $A$. By inspecting $\mathcal{IS}_Y(t)$, it results that no $Y\!\leftrightarrow\!B$ EPR exists. $Y$ can prepare this EPR pair and perform a local SWAP to \emph{establish} an entangled link between $A$ and $B$. Thus, the Planner builds a local PoA $D_Y=(A_Y,E_Y)$ reflecting the above, with actions such as:
\begin{align}
    \nonumber\{&\texttt{SYN}(Y\!-\!B),\ \texttt{GEN}(Y\!-\!B),\ \texttt{PURIFY}(Y\!-\!B),\\&
    \texttt{SWAP}(A\!-\!Y\!-\!B),\ \texttt{ACT\_FORWARD}(\widehat{B})\}.
\end{align}
Edges $E_Y$ encode logical and resource constraints, with optional parallel \texttt{GEN} if hinted: $\texttt{SYN}\!\rightarrow\!\texttt{GEN}^{\times k}\!\rightarrow\!\texttt{PURIFY}$ until $F\!\ge\!F_t$, then $\texttt{SWAP}\!\rightarrow\!\texttt{ACT\_FORWARD}$. 

\textit{Executor (mapping actions to MPs/MePs).} Feasibility and capability checks against $\mathcal{IS}_Y(t)$ select executable actions and bind them to MPs/MePs:
\begin{itemize}
    \item[-] \emph{$Y\!-\!B$ link preparation:} composes a meta-protocol $\mathrm{MeP}^{Y\!-\!B}_{\text{link}}$ as:$\ \textbf{MP\textsubscript{SYN}} \rightarrow \textbf{MP\textsubscript{GEN}}^{\times k} \rightarrow \textbf{MP\textsubscript{QP}}$ (local gates/measurements for purification), with classical feedback via \textbf{MP\textsubscript{SIG}} when required.
    \item[-] \emph{Local swapping:} composes the meta-protocol $\mathrm{MeP}_{\text{swap}}$ for performing the swapping as: \textbf{MP\textsubscript{QP}}$\rightarrow$ \textbf{MP\textsubscript{SIG}}, where \textbf{MP\textsubscript{QP}} at $Y$ is for the Bell-state measurement process and \textbf{MP\textsubscript{SIG}} is for disseminating the SWAP outcome to $B$, which is utilized not only to correct the shared entangled state, but also as a soft-state to update its quantum forwarding table $FT_q^B(t)$.
    \item[-] \emph{In-band forwarding:} implements \texttt{ACT\_FORWARD}(B) as \textbf{MP\textsubscript{FW}}$(\widehat{B}=B)$ for sending the updated meta-header to $B$, thereby enabling the BSM at A required by the teleportation protocol.
\end{itemize}
A schedule is produced that honors $E_Y$ and coherence windows; retries/backoffs remain encapsulated within $\mathrm{MeP}^{Y\!-\!B}_{\text{link}}$.

\textit{Engine (invoking MPs).} The Engine interfaces with the MP library and executes the scheduled MPs in order:
\begin{enumerate}
    \item[(i)] \textbf{MP\textsubscript{SYN}} on $Y\!-\!B$;
    \item[(ii)] repeated \textbf{MP\textsubscript{GEN}} on $Y\!-\!B$;
    \item[(iii)] purification via \textbf{MP\textsubscript{QP}} and \textbf{MP\textsubscript{SIG}} when required;
    \item[(iv)] \textbf{MP\textsubscript{QP}} for the SWAP at $Y$ and \textbf{MP\textsubscript{SIG}} to
        publish BSM outcome\footnote{As design strategy the BSM outcome is sent to only one of the node and not both.} to $B$;
    \item[(v)] \textbf{MP\textsubscript{FW}} to forward the meta-header to  $B$.
\end{enumerate}
In particular, after the entangled pair $Y\!-\!B$ reaches $(F_t,\tau)$ thresholds, \texttt{SWAP} commits by creating an entangled link between $A$ and $B$. At each action commit, the Engine appends a stamp $s$ in the meta-header. The commits at $Y$ -- an excerpt of which is reported in Tab.~\ref{tab:02} -- materialize as:
\begin{align}
    s_3&=(\texttt{LINK\_PR}(Y\!-\!B),\{Y,B\},T_3,\{\mathrm{e}_{YB}\},\textsf{ok},F_{YB},\tau_{YB}), \nonumber
 \\
    s_4&=(\texttt{SWAP}(A\!-\!Y\!-\!B),\{A,Y,B\},T_4,\{\mathrm{e}_{AB}\},\textsf{ok},F_{AB},\tau_{AB}), \nonumber \\
    s_5&=(\texttt{ACT\_FORWARD}(B),\{Y\},T_5,\varnothing,\textsf{ok},-). \nonumber
\end{align}

\vspace{3pt}
\subsubsection{Consumption across $A$ and $B$ (BSM at $A$, correction at $B$)}
Node $B$ receives $p=(H,\text{payload})$ from $Y$ with $H$ now including the \texttt{SWAP} stamp. By virtue of the SWAP at $Y$, nodes $A$ and $B$ hold their respective halves of the EPR pair in $FT_q^A(t)$ and $FT_q^B(t)$.

\textit{Kernel at $B$ (activate and notify $A$).} The Planner at $B$ adds the consumption and  forwarding actions:
\[
    \{\texttt{CONSUME\_TP}(A\!\rightarrow\!B),\ \texttt{ACT\_Deliver}\},
\]
with the Executor mapping \texttt{CONSUME\_TP} to \textbf{MP\textsubscript{QP}} at $B$ for the Pauli corrections and it schedules a \textbf{MP\textsubscript{SIG}} notification to $A$ (\texttt{ReadyForBSM}) to wake $A$’s hold. The Engine sends \textbf{MP\textsubscript{SIG}}.\texttt{ReadyForBSM}$\rightarrow A$ and awaits the two classical BSM bits from $A$.
 
\textit{Kernel at $A$ (perform BSM; meta-header remains at $B$).}
Upon \texttt{ReadyForBSM}, $A$’s Planner refines the local PoA with \(\{\texttt{BSM\_A}\}\). The Executor maps \texttt{BSM\_A} to \textbf{MP\textsubscript{QP}} on ($A$’s input qubit, $A$’s half of the $A$–$B$ ebit) and binds \textbf{MP\textsubscript{SIG}} to deliver the two outcomes to $B$. The Engine performs the BSM and sends the outcomes via \textbf{MP\textsubscript{SIG}}. No global stamp is appended at $A$ because $H$ resides at $B$; $A$ remains in \texttt{ACT\_HOLD} until completion.

\textit{Kernel at $B$ (apply correction; authoritative stamping).}
When the two BSM bits arrive, the kernel enables the correction. The Engine applies \textbf{MP\textsubscript{QP}} (Pauli) and appends:
\begin{align}
    s_6 &= \big(\texttt{CONSUME\_TP}(A\!\rightarrow\!B), \{A,B\}, T_6, \{\mathrm{e}_{AB}\}, \textsf{ok}, -\big), \nonumber \\
    s_{7}&= \big(\texttt{ACT\_DELIVER},\ \{B\},\ T_{7},\ \varnothing,\ \textsf{ok},\ -\big). \nonumber
\end{align}
It then emits \textbf{MP\textsubscript{SIG}}.\texttt{CompletionAck}$\rightarrow A$ so $A$ can close its local soft-state. The authoritative, in-band history is the monotone stamp sequence in $H$ held at $B$, which remains acyclic and monotone.

\textit{Closure at $A$ (local).} Upon receiving \texttt{CompletionAck}, $A$ transitions
\texttt{ACT\_HOLD}\,$\rightarrow$\,\texttt{ACT\_DELIVER} locally (soft-state).

\section{Discussion and Open Research Directions}
\label{sec:05}
\begin{table*}[t]
    \centering
    \renewcommand{\arraystretch}{1.5}
    \caption{Companion to Table~\ref{tab:01}: Dynamic Kernel remedies to layering limitations}
    \label{tab:03}
    \begin{tabular}{m{0.22\linewidth} m{0.35\linewidth} m{0.35\linewidth}}
        \hline
        \hline
        \textbf{Layering Features / Assumptions} & \textbf{Layered Design Consequence} & \textbf{Distributed-Kernel Remedy} \\
        \hline
        \multirow[t]{3}{0.9\linewidth}{\raggedright Vertical, static abstraction boundaries between functionalities}
            & Layers operate independently and inter-layer dependencies cannot be expressed. Local choices cannot safely constrain remote behavior.
            & No layering design. Each hop computes a local PoA and composes MPs/MePs at runtime. Resource/quality guards are encoded in PoA edges and are consulted via $\mathcal{IS}_v(t)$. Stamps certify progress. \\
        \cline{2-3}
            & Each layer expects its own local descriptors, causing inconsistency and overhead: state drifts as resources evolve. 
            & Single, authoritative in-band \emph{meta-header}. Only \emph{committed} actions append monotone stamps. No per-layer replicas, no drift. Orthogonal reads keep control quantum-native and non-perturbing. \\
        \cline{2-3}
            & No inherent adaptability to coherence budgets or ephemeral
resource availabilities.
            & PoA is rebuilt from $(\mathcal{I},\textsf{Stamps},\mathcal{IS}_v(t))$ at every hop; actions/MePs adapt to current resource availability and policy. \\
            \cline{2-3}
                & Poor expression of multi-hop, stateful behaviors: cross-layer sequences (e.g GEN$\rightarrow$PURIFY$\rightarrow$SWA) become brittle or duplicated; retries/backoffs leak across layers.
                & Kernel composes MPs into MePs that encode the sequence. Thus retries/backoffs stay inside MePs. Only terminal commits become stamps.\\
            \cline{2-3}
                & Functions/metadata duplicated to preserve layer autonomy, by hindering scalability.
                & Commit semantics and updates at action boundaries eliminate duplication. Header growth is bounded by \textit{committed} actions only. No speculative futures in-band. The stamp sequence is a compact, verifiable execution trace. This assures \textit{scalability}. \\
            \cline{2-3}
                & Technology evolution forces frequent interface changes. 
                & Planner is \emph{MP-agnostic}. The suite is invariant to MP/MeP boundary shifts: functionality can collapse or expand without changing header/stamp semantics. \\
        \hline
        \multirow[t]{2}{0.9\linewidth}{\raggedright Encapsulation as ordering primitive}
            & Encapsulation prescribes vertical order but cannot encode true dependencies (resource/quality guards) or cross-hop constraints.
            & Order merges (i) locally from the PoA DAG and schedule and (ii) globally from the stamp log, certifying commits and inducing an \emph{acyclic partial order} over action commits (no speculative edges exported). \\
        \hline
        \multirow[t]{3}{0.9\linewidth}{\raggedright Inherently classical control-plane}
            & Continuous, network-wide updates to maintain a global view scale poorly and incur latency. Thus assuring correctness is costly and fragile.
            & No global view required. Monotone stamps form a distributed commit log. A global DAG emerges from local decisions. Optional EDC hints prune conservative orderings. \\
        \cline{2-3}
            & 
            & The meta-header is an in-band control-field. Only terminal outcomes are stamped, by providing rollback-free failure handling. Classical signaling is  used strictly for essential needs (e.g., BSM bits). \\
        \hline
        \hline
    \end{tabular}
\end{table*}
This section summarizes some key properties of the proposed \textit{Distributed Dynamic Kernel} and positions it with respect to classical layering design. We distill how our proposal achieves modularity, adaptability, and scalability, and how processing order is enforced without fixed strata. A side-by-side mapping from layering conflicts to our remedies is provided in Table~\ref{tab:03}, as a companion to Table~\ref{tab:01}. 

\begin{itemize}
    \item{\textit{Decoupling of Mechanism and Implementation}:}
    A core tenet of the suite is the separation between \emph{mechanism} and \emph{implementation}. The \textit{Planner} is MP-agnostic, the \textit{Executor} maps actions to the available MP library, by binding concrete parameters, and the \textit{Engine} invokes MPs. As a result, correctness and semantics -- \emph{meta-header} and \emph{stamps} -- are invariant to where the MP/MeP boundary is drawn and to future technological shifts that may collapse/expand functionalities. This \emph{boundary invariance} preserves the suite across device generations, while enabling implementation evolution behind stable mechanisms.

    \item{\textit{Modularity via Opaque Primitives and Commit-Atomicity}:}
    Modularity is achieved by treating MPs as \emph{opaque} primitives with explicit pre/post-conditions and timing/coherence guards. The Executor composes MPs into MePs without depending on their internals. The Engine enforces \emph{commit-only} stamping: header/payload/$FT_q^v(t)$ updates become visible at action commit (or abort) boundaries. Retries/backoffs and intra-action loops are contained within MePs and never inflate the header. The stamp log thus becomes a compact, verifiable execution trace made only of \textit{facts that happened}, not speculative steps.

    \item{\textit{Adaptability and Scalability from Local PoA and In-Band History}:}
    Unlike fixed layering that prescribes a vertical order, the kernel computes a local PoA $D_v=(A_v,E_v)$ from $(\mathcal{I},\textsf{Stamps},\mathcal{IS}_v(t))$, adapting to current resource availability, coherence budgets, and policy. Progress is certified by \emph{monotone} action-commit stamps transported \emph{in-band} in an orthogonal subspace of the quantum packet. This eliminates per-layer duplication and removes the need for global control-plane refresh after each quantum step. Consequently, scalability follows from two principles: (i) \emph{local containment} of control-flow and failures, and (ii) \emph{minimal growth} of the header, which increases only with committed actions.

    \item{\textit{Order by Certification Not by Prescription}:}
    In classical layering, encapsulation and fixed interfaces \emph{prescribe} a vertical processing order. In the proposed suite, processing order \emph{emerges} from three complementary mechanisms:
    \begin{enumerate}
        \item{\textit{Local enforcement.}} The Planner $D_v(t)=(A_v(t),E_v(t))$ and the Executor schedule enforce intra-node sequencing, by honoring dependency edges and resource/quality guards.
        \item{\textit{Commit boundaries.}} The Engine action boundaries (commit/abort) are the only points where header/payload/$FT_q(t)$ changes take effect.
        \item{\textit{Global propagation.}} Monotone action-commit stamps carried in-band are the authoritative, minimal history that the next hop must respect. They do not prescribe future steps, rather they induce an emergent, network-wide partial order $D^*(t)$ over what has actually committed.
    \end{enumerate}
    As a consequence, \textit{within a node} order is given by $E_v$ and the scheduled MPs, while \textit{across nodes} order is the stamped commit-to-commit ``happens-before''. Optionally, EDC hints can prune view-induced serializations. In a nutshell, while encapsulation enforces order by \textit{prescription} in fixed layering-based stacks, our design enforces order by \textit{certification}: local scheduling decides what may run, stamps certify what did run, and subsequent nodes must plan in ways consistent with that certified history.
\end{itemize}

From the above, it emerges that the design: (i) decouples mechanism from implementation via MP-agnostic planning and commit semantics; (ii) preserves modularity through opaque MPs and MePs; (iii) achieves adaptability by computing PoAs from local state and in-band history; and (iv) scales by avoiding global synchronization and by bounding header growth to committed actions. 

\subsection{Open Research Directions}
\label{sec:5.1}

While this work establishes a quantum-native organizational principle for the Quantum Internet, several research directions remain open for the community to explore and investigate.

\paragraph*{Physical realization of the meta-header}
In our proposal, the \textit{meta-header} is an architectural control abstraction and is intentionally agnostic to any specific physical encoding. Several realizable options exist \cite{CalDavHan-25}. For instance, orthogonality between the meta-header and the payload subspaces (required for non-perturbing access) can be implemented either (i) by exploiting distinct degrees of freedom -- e.g., time-bin, frequency-bin, spatial-mode, or polarization -- of optical photons or (ii) by encoding control and payload into mutually orthogonal subspaces within the same degree of freedom. A systematic comparison of these alternatives is an interesting research direction. Future work could quantify qubit/mode overhead, control-field lifetime, error accumulation associated with specific encoding choices and constraints. This thereby will establish concrete trade-offs between architectural generality and encoding-specific choices.

\paragraph*{Quantitative Cost Models}
Complementary to the above physical-realization direction, a second open direction is the development of quantitative cost models. Specifically, while as indicated in Sec.~\ref{sec:3.3} the meta-header growth is architecturally bounded and semantically constrained by the number of committed actions required by a service intent, translating this bound into platform-specific overheads -- e.g., number of qubits/modes per stamp, encoding-dependent error rates, and read/write latency -- for different service classes requires careful investigation in its own right. Such analysis would yield explicit complexity relations linking service semantics -- number and type of action commits -- to the physical footprint of the resulting stamp sequence, and it would enable principled comparisons across alternative encoding and implementation choices. Importantly, any such cost model should respect the architectural constraints of this work: meta-header and payload must remain operationally decoupled, and commit-atomic semantics must be preserved.

\paragraph*{Simulation of the Dynamic Kernel}
An important and very interesting open-research direction is the systematic evaluation of the proposed organizational model on quantum network simulators. In this perspective, we have already initiated the development of a quantum network simulator explicitly conceived to natively support Dynamic-Kernel–based protocol composition, with preliminary results reported in~\cite{PeaMazCal-26}. Extending this simulator to incorporate the full kernel pipeline -- commit semantics, in-band meta-header evolution, MP/MeP scheduling, resource-aware PoA construction, and stochastic failure models -- will enable quantitative assessment of performance, scalability, and robustness under realistic network dynamics. Such a simulator provides a controlled environment to explore alternative encoding choices and derive service-dependent cost models consistent with the commit-driven design.

\paragraph*{Experimental mapping on near-term testbeds}
A complementary research direction concerns the experimental instantiation of the Dynamic Kernel abstractions onto near-term quantum-network testbeds. Our ongoing effort on an in-house photonic-entanglement Quantum Internet testbed aims at translating PoA construction, MP/MeP composition, and commit-based meta-header processing into executable components on photonic hardware. Experimental validation will provide empirical insight into coherence-time budgeting, classical side-channel reliability, end-to-end latency accumulation, and the operational behavior of distributed commit semantics in non-ideal regimes. Together, simulation and testbed investigations constitute a substantial research program enabled by the architectural foundation established in this work. And we hope the broader community will engage with these efforts.

\paragraph*{Standardization and Industrial Relevance}
An open research direction with direct industrial impact is how to translate the proposed organizational principle into \textit{standardizable} abstractions \cite{CacCalIll-25,rfc9340}. Moving from research prototypes to production-scale Quantum Internet infrastructures will require architectural primitives that remain stable under rapid hardware evolution and heterogeneous vendor implementations. In this perspective, the separation between \textit{mechanism} (MP semantics), \textit{composition} (MePs), and \textit{certification} (stamps and commit semantics) positions the Dynamic Kernel as a candidate foundation for future standardization efforts. Key questions include, but are not limited to: (i) identifying a minimal and extensible meta-header format; (ii) determining which stamp fields are mandatory versus optional for each service class; and (iii) defining verifiable conformance criteria for commit semantics and in-band control invariants. Addressing these questions would enable interoperability across platforms and vendors, foster ecosystem-level convergence, and ultimately accelerate industrial adoption and technology transfer.

\medskip

Collectively, these directions illustrate that the present work provides a foundational architectural layer. Its value lies not only in replacing static layering, but in defining a coherent design space within which scalable, interoperable, and industrially viable quantum networking systems can be systematically engineered.

\section{Conclusion}
\label{sec:06}

Static layering, while historically successful in classical networking, is structurally misaligned with the non-local and stateful nature of entanglement. In a quantum network, local actions at a node constrain remote possibilities, preventing service progression from being cleanly decomposed into vertically isolated strata. 
For this, we introduced a \textit{quantum-native organizational principle} based on \textit{dynamic composition}. Within the designed protocol suite, each node runs a \textit{Dynamic Kernel} that constructs a local PoA from the service intent, accumulated commit stamps, and its internal state, and then executes the currently feasible actions by composing micro-protocols (MPs) into meta-protocols (MePs). Service progress is certified through in-band, monotone action-commit records carried in the meta-header, which collectively induce a distributed DAG of committed actions without requiring global knowledge or network-wide synchronization. The proposed principle enforces order by certification rather than prescription: local scheduling determines what may run, commit semantics determine what did run, and subsequent planning must respect that certified history. Correctness is entirely local and in-band, thereby preserving scalability, while performance can optionally be enhanced via advisory control-plane hints without changing protocol semantics. This commit-driven, state-aware framework constitutes a necessary shift in protocol organization for the Quantum Internet. It provides a hardware-agnostic -- yet operationally precise -- foundation on which simulators, experimental testbeds, and ultimately interoperable standards can be built across heterogeneous platforms. Hardware-agnostic organizational principles are essential to withstand technology evolution. By separating mechanism from implementation, and by remaining encoding-agnostic while explicitly entanglement-state-aware, the proposed Dynamic Kernel offers a stable architectural framework for a scalable and market-ready Quantum Internet.

\bibliographystyle{ieeetr}
\bibliography{biblio.bib,biblio2.bib}

@article{VanSatBen-21,
    doi = {10.1109/QCE53715.2022.00055},
    year = {2022},
    month = {sep},
    pages = {341-352},
    title = {A Quantum Internet Architecture},
    author = {Van Meter, R. and others},
    address = {Los Alamitos, CA, USA},
    journal = {IEEE QCE22},
    publisher = {IEEE Computer Society}}

@article{KriClaFal-07,
    doi = {10.1145/1273445.1273450},
    year = {2007},
    month = {jul},
    pages = {41–52},
    title = {On Compact Routing for the Internet},
    author = {Krioukov, Dmitri and claffy, k c and Fall, Kevin and Brady, Arthur},
    number = {3},
    volume = {37},
    journal = {SIGCOMM Comput. Commun. Rev.},
    publisher = {Association for Computing Machinery},
    issue_date = {July 2007}}

@article{Kim-08,
    year = {2008},
    pages = {1023--1030},
    title = {The quantum internet},
    author = {Kimble, H Jeff},
    number = {7198},
    volume = {453},
    journal = {Nature},
    publisher = {Nature Publishing Group}}

@article{VanLadMun-08,
    year = {2008},
    pages = {1002--1013},
    title = {System design for a long-line quantum repeater},
    author = {Van Meter, Rodney and Ladd, Thaddeus D and Munro, William J and Nemoto, Kae},
    number = {3},
    volume = {17},
    journal = {IEEE/ACM Transactions on Networking},
    publisher = {IEEE}}

@article{VanTouHor-11,
    issn = {1349-8614},
    year = {2011},
    issue = {8},
    pages = {65--79},
    title = {Recursive quantum repeater networks},
    author = {Van Meter, Rodney and Touch, Joe and Horsman, Clare},
    journal = {NII Journal},
    publisher = {National Institute of Informatics}}

@article{VanTou-13,
    year = {2013},
    pages = {64-71},
    title = {Designing quantum repeater networks},
    author = {Meter, Rodney Van and Touch, Joe},
    number = {8},
    volume = {51},
    journal = {IEEE Commun. Mag.}}

@article{DurLamHeu-17,
    year = {2017},
    pages = {043001},
    title = {Towards a quantum internet},
    author = {D{\"u}r, Wolfgang and Lamprecht, Raphael and Heusler, Stefan},
    number = {4},
    volume = {38},
    journal = {European Journal of Physics},
    publisher = {IOP Publishing}}

@article{MatDurVan-19,
    year = {2019},
    issue = {5},
    month = {Nov},
    pages = {052320},
    title = {Quantum link bootstrapping using a RuleSet-based communication protocol},
    author = {Matsuo, Takaaki and Durand, Cl\'ement and Van Meter, Rodney},
    volume = {100},
    journal = {Phys. Rev. A},
    numpages = {13},
    publisher = {American Physical Society}}

@inproceedings{KozDahWeh-20,
    year = {2020},
    pages = {1--16},
    title = {Designing a quantum network protocol},
    author = {Kozlowski, Wojciech and Dahlberg, Axel and Wehner, Stephanie},
    booktitle = {Proc. of the 16th International Conference on emerging Networking EXperiments and Technologies}}

@article{CacCalVan-20,
    note = {invited paper},
    year = {2020},
    pages = {3808--3833},
    title = {When entanglement meets classical communications: Quantum teleportation for the quantum internet},
    author = {Cacciapuoti, Angela Sara and others},
    number = {6},
    volume = {68},
    journal = {IEEE Trans. on Communications},
    publisher = {IEEE}}

@article{PirDur-19,
    year = {2019},
    month = {mar},
    pages = {033003},
    title = {A quantum network stack and protocols for reliable entanglement-based networks},
    author = {A Pirker and W Dür},
    number = {3},
    volume = {21},
    journal = {New Journal of Physics},
    publisher = {{IOP} Publishing}}

@inproceedings{Weh-19,
    isbn = {9781450359566},
    year = {2019},
    pages = {159–173},
    title = {A Link Layer Protocol for Quantum Networks},
    author = {Dahlberg, Axel and others},
    numpages = {15},
    booktitle = {Proc. of ACM SIGCOMM '19}}

@inproceedings{KozWeh-19,
    year = {2019},
    pages = {1--7},
    title = {Towards large-scale quantum networks},
    author = {Kozlowski, Wojciech and Wehner, Stephanie},
    booktitle = {Proc. of ACM NANOCOM '19}}

@article{RamPirDur-21,
    year = {2021},
    title = {Genuine quantum networks with superposed tasks and addressing},
    author = {Miguel-Ramiro, J. and others},
    journal = {npj Quantum Inf}}

@article{PomDonWeh-21,
    year = {2022},
    title = {Experimental demonstration of entanglement delivery using a quantum network stack},
    author = {Pompili, Matteo and others},
    journal = {npj Quantum Information}}

@article{IllCalMan-22,
    doi = {https://doi.org/10.1016/j.comnet.2022.109092},
    url = {https://www.sciencedirect.com/science/article/pii/S1389128622002250},
    issn = {1389-1286},
    year = {2022},
    pages = {109092},
    title = {Quantum Internet protocol stack: A comprehensive survey},
    author = {Jessica Illiano and Marcello Caleffi and Antonio Manzalini and Angela Sara Cacciapuoti},
    volume = {213},
    journal = {Computer Networks}}

@article{CacCalTaf-20,
    doi = {10.1109/MNET.001.1900092},
    year = {2020},
    pages = {137-143},
    title = {Quantum Internet: Networking Challenges in Distributed Quantum Computing},
    author = {Cacciapuoti, Angela Sara and Caleffi, Marcello and Tafuri, Francesco and Cataliotti, Francesco Saverio and Gherardini, Stefano and Bianchi, Giuseppe},
    number = {1},
    volume = {34},
    journal = {IEEE Network}}

@article{CacIllCal-23,
    year = {2023},
    title = {Quantum Internet Addressing},
    author = {Cacciapuoti, Angela Sara and Illiano, Jessica and Caleffi, Marcello},
    journal = {IEEE Network},
    publisher = {IEEE}}

@article{MazCalCac-25,
    doi = {10.1109/TNSE.2024.3520856},
    year = {2025},
    pages = {1-18},
    title = {{Intra-QLAN Connectivity via Graph States: Beyond the Physical Topology}},
    author = {Mazza, Francesco and Caleffi, Marcello and Cacciapuoti, Angela Sara},
    journal = {IEEE Transactions on Network Science and Engineering}}

@article{DiAQiMil22,
    title = {Packet switching in quantum networks: A path to the quantum Internet},
    author = {DiAdamo, Stephen and Qi, Bing and Miller, Glen and Kompella, Ramana and Shabani, Alireza},
    journal = {Phys. Rev. Res.},
    volume = {4},
    year = {2022},
    month = {Oct}
}

@ARTICLE{YooSinKum24,
  author={Ben Yoo, S. J. and Singh, Sandeep Kumar and On, Mehmet Berkay and Gül, Gamze and Kanter, Gregory S. and Proietti, Roberto and Kumar, Prem},
  journal={IEEE Communications Magazine}, 
  title={Quantum Wrapper Networking}, 
  year={2024},
  volume={62}
}

@ARTICLE{VisHolDia24,
    author={Vista, Francesco and Holme, Daniel and DiAdamo, Stephen},
    journal={IEEE Communications Magazine}, 
    title={Quantum Backbone Networks for Hybrid Quantum Dataframe Transmission}, 
    year={2024},
    volume={62}
}

@article{CalDavHan-25,
  author={Caleffi, Marcello and d’Avossa, Laura and Han, Xu and Sara Cacciapuoti, Angela},
  journal={IEEE Communications Surveys \& Tutorials}, 
  title={Quantum Transduction: Enabling Quantum Networking}, 
  year={2026},
  volume={28},
  doi={10.1109/COMST.2025.3631150}}

@book{VanMet-14, 
	author = {Van Meter, Rodney}, 
	title = {Quantum Networking}, 
	year = {2014}, isbn = {1848215371}, 
	publisher = {Wiley-IEEE Press}, 
	edition = {1st}
}

@article{SimCalIll-23,
    title = {Universal quantum computation via superposed orders of single-qubit gates},
    author = {Simonov, Kyrylo and Caleffi, Marcello and Illiano, Jessica and Romero, Jacquiline and Cacciapuoti, Angela Sara},
    journal = {Phys. Rev. Res.},
    year = {2025},
    month = {Dec},
    publisher = {American Physical Society},
    doi = {10.1103/6xwr-9b4y},
    url = {https://link.aps.org/doi/10.1103/6xwr-9b4y}
}

@article{CheIllCac-24,
  author={Chen, Si-Yi and Illiano, Jessica and Cacciapuoti, Angela Sara and Caleffi, Marcello},
  journal={IEEE Open Journal of the Communications Society}, 
  title={Entanglement-Based Artificial Topology: Neighboring Remote Network Nodes}, 
  year={2025},
  volume={6},
  number={},
  pages={2220-2238},
  doi={10.1109/OJCOMS.2025.3554052}}

@misc{rfc9583,
    series =    {Request for Comments},
    number =    9583,
    howpublished =  {RFC 9583},
    publisher = {RFC Editor},
    author =    {Chonggang Wang and Akbar Rahman and Ruidong Li and Melchior Aelmans and Kaushik Chakraborty},
    title =     {{Application Scenarios for the Quantum Internet}},
    pagetotal = 26,
    year =      2024,
    month =     jun
}

@misc{rfc9340,
    series =    {Request for Comments},
    number =    9340,
    howpublished =  {RFC 9340},
    publisher = {RFC Editor},
    doi =       {10.17487/RFC9340},
    url =       {https://www.rfc-editor.org/info/rfc9340},
    author =    {Wojciech Kozlowski and Stephanie Wehner and Rodney Van Meter and Bruno Rijsman and Angela Sara Cacciapuoti and Marcello Caleffi and Shota Nagayama},
    title =     {{Architectural Principles for a Quantum Internet}},
    pagetotal = 37,
    year =      2023,
    month =     mar
}

@article{CalCac-25,
  title={{Quantum Internet Architecture: unlocking Quantum-Native Routing via Quantum Addressing}},
  author={Caleffi, Marcello and Cacciapuoti, Angela Sara},
  journal={IEEE Transactions on Communications},
  doi = {10.1109/TCOMM.2025.3650397},
  year={2026},
  note = {invited paper}
}

@misc{PirMunDur-25,
      title={A resource-centric, task-based approach to quantum network control}, 
      author={Alexander Pirker and Belen Munoz and Wolfgang Dür},
      year={2025},
      eprint={2507.12030},
      archivePrefix={arXiv}
}

@ARTICLE{KreRamVer-14,
  author={{{Kreutz, Diego and Ramos, Fernando M. V. and Veríssimo, Paulo Esteves and Rothenberg, Christian Esteve and Azodolmolky, Siamak and Uhlig, Steve}}},
  journal={Proceedings of the IEEE}, 
  title={Software-Defined Networking: A Comprehensive Survey}, 
  year={2015},
  volume={103}}

@techreport{CacCalIll-25,
    number =    {draft-cacciapuoti-qirg-quantum-native-architecture-00},
    type =      {Internet-Draft},
    institution =   {Internet Engineering Task Force},
    publisher = {Internet Engineering Task Force},
    note =      {Work in Progress},
    url =       {https://datatracker.ietf.org/doc/draft-cacciapuoti-qirg-quantum-native-architecture/00/},
    author =    {Angela Sara Cacciapuoti and Marcello Caleffi and Jessica Illiano and C. De Risi},
    title =     {{Quantum-Native Architectural Tenets and Philosophy for the Quantum Internet}},
    pagetotal = 14,
    year =      2025,
    month =     nov,
    day =       14
}

@article{BraFabHan-03, 
author = {Braden, Robert and Faber, Ted and Handley, Mark}, title = {From protocol stack to protocol heap: role-based architecture}, year = {2003}, issue_date = {January 2003}, publisher = {Association for Computing Machinery}, address = {New York, NY, USA}, volume = {33}, number = {1}, issn = {0146-4833}, url = {https://doi.org/10.1145/774763.774765}, doi = {10.1145/774763.774765}, journal = {SIGCOMM Comput. Commun. Rev.}, month = jan, pages = {17–22}, numpages = {6} }

@ARTICLE{GazPatAlo-10,
  author={Gazis, Vangelis and Patouni, Eleni and Alonistioti, Nancy and Merakos, Lazaros},
  journal={IEEE Communications Surveys \& Tutorials}, 
  title={A survey of dynamically adaptable protocol stacks}, 
  year={2010},
  volume={12},
  number={1},
  pages={3-23},
  doi={10.1109/SURV.2010.020110.00034}}

@book{Day-08,
  author    = {John Day},
  title     = {{Patterns in Network Architecture: A Return to Fundamentals}},
  year      = {2008},
  publisher = {Prentice Hall}
}

@ARTICLE{LiXueLi-23,
  author={Li, Zhonghui and Xue, Kaiping and Li, Jian and Chen, Lutong and Li, Ruidong and Wang, Zhaoying and Yu, Nenghai and Wei, David S. L. and Sun, Qibin and Lu, Jun},
  journal={IEEE Communications Surveys \& Tutorials}, 
  title={Entanglement-Assisted Quantum Networks: Mechanics, Enabling Technologies, Challenges, and Research Directions}, 
  year={2023},
  volume={25},
  number={4}}

@ARTICLE{LiuAllCai-22,
  author={Liu, Maoli and Allcock, Jonathan and Cai, Kechao and Zhang, Shengyu and Lui, John C.S.},
  journal={IEEE Network}, 
  title={Quantum Networks with Multiple Service Providers: Transport Layer Protocols and Research Opportunities}, 
  year={2022},
  volume={36},
  number={5}}

@ARTICLE{WanRah-22,
  author={Wang, Chonggang and Rahman, Akbar},
  journal={IEEE Wireless Communications}, 
  title={Quantum-Enabled 6G Wireless Networks: Opportunities and Challenges}, 
  year={2022},
  volume={29},
  number={1}}

@ARTICLE{WuHuLi-24,
  author={Wu, Jindi and Hu, Tianjie and Li, Qun},
  journal={IEEE Network}, 
  title={Q-ID: Lightweight Quantum Network Server Identification Through Fingerprinting}, 
  year={2024},
  volume={38},
  number={5}}

@article{PeaMazCal-26,
  title={{Q2NS: A Modular Framework for Quantum Network Simulation in ns-3}},
  author={Pearson, Adam and Mazza, Francesco and Caleffi, Marcello and Cacciapuoti, Angela Sara},
  journal={to appear in Proc. of IEEE International Conference on Quantum Communications, Networking, and Computing (IEEE QCNC)},
  year={2026},
  note = {invited paper}
}

\end{document}